\documentclass{article}

    \usepackage[final]{neurips_2021}

\usepackage{graphicx}
\graphicspath{ {../images/} }
\usepackage[utf8]{inputenc} 
\usepackage[T1]{fontenc}    
\usepackage{hyperref}       
\usepackage{url}            
\usepackage{booktabs}       
\usepackage{amsfonts}       
\usepackage{nicefrac}       
\usepackage{microtype}      
\usepackage{xcolor}         
\usepackage{comment}
\usepackage{algorithm}
\usepackage{algorithmic}
\usepackage{xspace}

\usepackage{booktabs}
\usepackage{array}
\newcolumntype{L}{>{\centering\arraybackslash}m{1.7cm}}

\usepackage{upgreek}
\usepackage{amsmath}
\usepackage{amssymb}
\usepackage{subcaption}
\usepackage{multirow}
\usepackage[utf8]{inputenc}
\usepackage{amsthm}

\usepackage{enumitem}

\usepackage{algorithm}
\usepackage{algorithmic}
\usepackage{multicol}

\DeclareMathOperator{\E}{\mathbb{E}}
\DeclareMathOperator*{\argmax}{arg\,max}

\newcommand{\todopb}[1]{\textcolor{red}{#1}\xspace}

\newcommand{\ipdistill}{$GameDistill$\xspace}
\renewcommand\cite{\citep}
\newtheorem{lemma}{Lemma}

\newcommand{\abhishek}[1]{}

\title{Status-quo policy gradient in Multi-Agent Reinforcement Learning}

\author{%
 Pinkesh Badjatiya\footnotemark[1]\thanks{Work done while at Media and Data Science Research Labs, Adobe}\\
  Microsoft, India\\
  \texttt{pbadjatiya@microsoft.com}
  \And
 Mausoom Sarkar\\
  Media and Data Science Research Lab, Adobe
  \And
  Nikaash Puri \\
  Media and Data Science Research Lab, Adobe
 \And
 Jayakumar Subramanian\\
  Media and Data Science Research Lab, Adobe
  \And
  Abhishek Sinha\footnotemark[1]\\
  Waymo\\
  \texttt{a7b23@stanford.edu}
  \And
  Siddharth Singh\footnotemark[2]\thanks{Work done during the internship at Adobe}\\
  University of Maryland\\
  \texttt{siddharth9820@gmail.com}
  \And
  Balaji Krishnamurthy\\
  Media and Data Science Research Lab, Adobe \\\\
  \texttt{\{msarkar,jasubram,nikpuri,kbalaji\}@adobe.com}
}

\begin{document}
\maketitle
\begin{abstract}
    Individual rationality, which involves maximizing expected individual return, does not always lead to high-utility individual or group outcomes in multi-agent problems. 
    For instance, in multi-agent social dilemmas, Reinforcement Learning (RL) agents trained to maximize individual rewards converge to a low-utility mutually harmful equilibrium. 
    In contrast, humans evolve useful strategies in such social dilemmas.
    Inspired by ideas from human psychology that attribute this behavior to the status-quo bias, we present a status-quo loss ($SQLoss$) and the corresponding policy gradient algorithm that incorporates this bias in an RL agent.
    We demonstrate that agents trained with $SQLoss$ learn high-utility policies in several social dilemma matrix games (Prisoner's Dilemma, Stag Hunt matrix variant, Chicken Game).
    We show how $SQLoss$ outperforms existing state-of-the-art methods to obtain high-utility policies in visual input non-matrix games (Coin Game and Stag Hunt visual input variant) using pre-trained cooperation and defection oracles. 
    Finally, we show that $SQLoss$ extends to a 4-agent setting by demonstrating the emergence of cooperative behavior in the popular Braess' paradox.

\end{abstract}
\section{Introduction}
\label{sec:introduction}

In sequential social dilemmas, individually rational behavior can lead to outcomes that are sub-optimal for each individual in the group \citet{Hardin1968,ostrom:1990,Ostrom278,Dietz1907}. 
Current state-of-the-art Multi-Agent Deep Reinforcement Learning (MARL) methods that train agents independently lead to agents that play individualistically, thereby receiving poor rewards, even in simple social dilemmas \cite{foerster2018learning, Peysakhovich1707.01068}.

To illustrate why it is challenging to learn optimal policies in such dilemmas, we consider the Coin Game, which was first explored in ~\cite{foerster2018learning}.
\abhishek{This line gives the feeling that you are the first to explore Coin game in social dima settings. However, I think that this is an already explored game right? } 
Each agent can play either selfishly (pick all coins) or cooperatively (pick only coins of its color). 
Regardless of the other agent's behavior, the individually rational choice for an agent is to play selfishly, either to minimize losses (avoid being exploited) or to maximize gains (exploit the other agent).
However, when both agents behave rationally, they try to pick all coins and achieve an average long term reward of $-0.5$. 
In contrast, if both cooperate, then the average long term reward for each agent is $0.5$. 
Therefore, when agents cooperate, they are both better off. \abhishek{Too deep an explanation of the game, can shorten it }
Training Deep RL agents independently in the Coin Game using state-of-the-art methods leads to mutually harmful selfish behavior (Section~\ref{sec:approach:selfish-learner}).

The problem of how independently learning agents develop optimal behavior in social dilemmas has been studied by researchers through human studies and simulation models~\cite{Fudenberg, Edward:1984, Fudenberg:1984, Kamada:2010, Abreu:1990}. 
A large body of work has looked at the mechanism of evolution of cooperation through reciprocal behaviour and indirect reciprocity~\cite{reciprocal1971,reciprocal1984,reciprocal1992,reciprocal1993,in_reciprocal1998}, through variants of reinforcement using aspiration~\cite{reinforce_variant}, attitude~\cite{NonRL_attitude} or multi-agent reinforcement learning~\cite{Sandholm1996MultiagentRL,Wunder:2010}, and under specific conditions~\cite{R_plus_S_g_2P} using different learning rates \cite{deCote:2006} similar to WoLF~\cite{WOLF2002} as well as using embedded emotion~\cite{Emotional_Multiagent}, social networks~\cite{Ohtsuki2006,Santos:2006}.
\abhishek{This looks like a paragraph to go in the related works? Not really needed here. Also can change the citation style to display numbers instead of author names. Can just mention a few of the related works here, and remaining in the related work section?}
\begin{figure*}[t!]
    \centering
    \includegraphics[width=\linewidth]{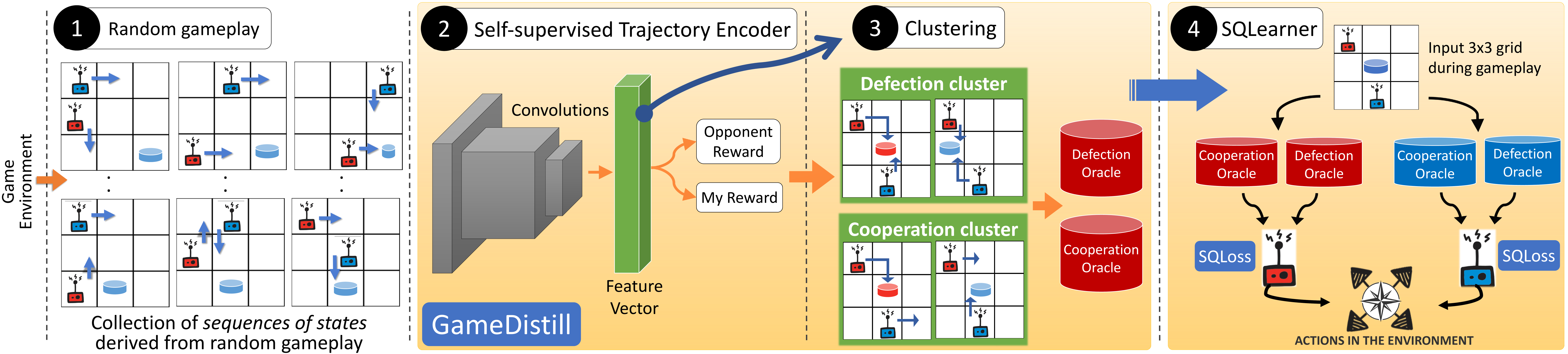}
    \caption{High-level architecture illustrated using coin game. Each agent runs \ipdistill by performing steps $(1), (2), (3)$ individually to obtain two oracles per agent. During game-play$(4)$, each agent (with $SQLoss$) takes either the action suggested by the cooperation or the defection oracle}
    \label{fig:high_level_approach}
\end{figure*}

However, these approaches do not directly apply to Deep RL agents~\cite{Leibo:2017}. 
Recent work in this direction~\cite{kleiman2016coordinate,Julien:2017,consequentialist18} focuses on letting agents learn strategies through interactions with other agents in a multi-agent setting. 
\citet{Leibo:2017} defines the problem of social dilemmas in the Deep RL framework and analyzes the outcomes of a fruit-gathering game~\cite{Julien:2017}. 
They vary the abundance of resources and the cost of conflict in the fruit environment to generate degrees of cooperation between agents. 
\citet{Hughes:2018} defines an intrinsic reward (inequality aversion) that attempts to reduce the difference in obtained rewards between agents. 
The agents are designed to have an aversion to both advantageous (guilt) and disadvantageous (unfairness) reward allocation. 
This handcrafting of loss with mutual fairness develops cooperation, but it leaves the agent vulnerable to exploitation. 
LOLA~\cite{foerster2018learning} uses opponent awareness to achieve high cooperation levels in the Coin Game and the Iterated Prisoner's Dilemma game. 
However, the LOLA agent assumes access to the other agent's network architecture, observations, and learning algorithms.  
This access level is analogous to getting complete access to the other agent's private information and therefore devising a strategy with full knowledge of how they are going to play.  
\citet{Wang:2019} proposes an evolutionary Deep RL setup to learn cooperation. 
They define an intrinsic reward that is based on features generated from the agent's past and future rewards, and this reward is shared with other agents. 
They use evolution to maximize the sum of rewards among the agents and thus learn cooperative behavior. 
However, sharing rewards in this indirect way enforces cooperation rather than evolving it through independently learning agents. 
\abhishek{Again this paragraph is just literature review. No need to mention all approaches, and even if you are mentioning it should be max 1 line for each. I think only LOLA can be mentioned here}

Interestingly, humans develop individual and socially optimal strategies in such social dilemmas without sharing rewards or having access to private information. 
Several ideas in human psychology~\cite{samuelson1988status, kahneman1991anomalies, kahneman2011thinking, thaler2009nudge} have attributed this cooperative behavior to the status-quo bias~\cite{guney2018costly}.
The status-quo bias is a decision-making bias that encourages humans to stick to the status-quo unless doing so causes significant harm. 
This preference in humans for sticking to `comfortable' states rather than seeking short-term exploitative gains is one explanation for emergent cooperative behavior in human groups.
Inspired by this idea, we present the status-quo loss ($SQLoss$) and the corresponding status-quo policy gradient formulation for RL.  
Agents trained with $SQLoss$ learn optimal policies in multi-agent social dilemmas without sharing rewards, gradients, or using a communication channel.
Intuitively, $SQLoss$ encourages an agent to stick to past actions provided these actions did not cause significant harm.
Therefore, mutually cooperating agents stick to cooperation since the status-quo yields higher individual reward, while unilateral defection by any agent leads to the other agent also switching to defection since the status-quo is very harmful for the exploited agent. 
Subsequently, for each agent, the short-term reward of exploitation is overcome by the long-term cost of mutual defection, and agents gradually switch to cooperation. 

To apply $SQLoss$ to games where a sequence of non-trivial actions determines cooperation and defection, we present \ipdistill, an algorithm that reduces a dynamic game with visual input to a matrix game. 
\ipdistill uses self-supervision and clustering to extract distinct policies from a sequential social dilemma game automatically. 

Our key contributions can be summarised as:
\begin{enumerate}[topsep=0pt,itemsep=1ex,partopsep=1ex,parsep=1ex]
    \item We introduce a \textbf{Status-Quo} loss ($SQLoss$, Sec.~\ref{sec:approach:SQLoss}) and an associated policy gradient-based algorithm to learn optimal behavior in a decentralized manner for agents playing iterated matrix games where agents can choose between a cooperative and selfish policy at each step. 
    We empirically demonstrate that agents trained with $SQLoss$ learn optimal behavior in several social dilemma iterated matrix games (Sec. \ref{sec:results}).
    \item We extend $SQLoss$ to social dilemma game with visual observations (Sec.~\ref{sec:GameDistill}) using \ipdistill. We empirically demonstrate that \ipdistill extracts cooperative and selfish policies for the Coin Game and Stag Hunt (Sec.~\ref{sec:results:visual-input-gamedistll-and-sqloss}). We further show that incorporating \ipdistill in LOLA accelerates learning and achieves higher cooperation.
    \item We also demonstrate that $SQLoss$ extends in a straight forward manner to games with more than two agents due to its ego-centric nature (Sec.~\ref{sec:beyond_2_players}).
\end{enumerate}
\section{Approach}
\label{sec:approach}

\subsection{Social Dilemmas modeled as Iterated Matrix Games}
To remain consistent with previous work, we adopt the notations from \citet{foerster2018learning}.
We model social dilemmas as general-sum Markov (simultaneous move) games. A multi-agent Markov game is specified by $G = \langle$$S$, $A$, $U$, $P$, $r$, $n$, $\gamma$$\rangle$. $S$ denotes the state space of the game. $n$ denotes the number of agents playing the game.
At each step of the game, each agent $a \in A$, selects an action $u^a \in U$. $\Vec{u}$ denotes the joint action vector that represents the simultaneous actions of all agents. The joint action $\Vec{u}$ changes the state of the game from $s$ to $s'$ according to the state transition function $P(s'|\Vec{u},s): S \times \textbf{U} \times S \rightarrow  [0, 1]$. At the end of each step, each agent $a$ gets a reward according to the reward function $r^{a}(s, \Vec{u}): S \times \textbf{U} \rightarrow \mathbb{R}$. The reward obtained by an agent at each step is a function of the actions played by all agents. For an agent $a$, the discounted future return from time $t$ is defined as $R_{t}^{a} = \sum_{l=0}^{\infty} \gamma^{l} r_{t+l}^{a}$, where $\gamma \in [0, 1)$ is the discount factor. Each agent independently attempts to maximize its expected discounted return. 

Matrix games are the special case of two-player perfectly observable Markov games~\cite{foerster2018learning}.  Table~\ref{tab:payoff_matrix} shows examples of matrix games that represent social dilemmas. 
Consider the Prisoner's Dilemma game in Table~\ref{tab:payoff_matrix_ipd}. Each agent can either cooperate (\textit{C}) or defect (\textit{D}). Playing \textit{D} is the rational choice for an agent, regardless of whether the other agent plays \textit{C} or \textit{D}. Therefore, if both agents play rationally, they each receive a reward of $-2$. However, if each agent plays \textit{C}, then it will obtain a reward of $-1$. This fact that individually rational behavior leads to a sub-optimal group (and individual) outcome highlights the dilemma. 

In Infinitely Iterated Matrix Games, agents repeatedly play a particular matrix game against each other. In each iteration of the game, each agent has access to actions played by both agents in the previous iteration. Therefore, the state input to an RL agent consists of both agents' actions in the previous iteration of the game.
We adopt this state formulation as is typically done in such games \cite{press2012iterated, foerster2018learning}.
The infinitely iterated variations of the matrix games in Table~\ref{tab:payoff_matrix} represent sequential social dilemmas. 
We refer to infinitely iterated matrix games as iterated matrix games in subsequent sections for convenience. 

\subsection{Learning Policies in Iterated Matrix Games: The Selfish Learner}
\label{sec:approach:selfish-learner}
The standard method to model agents in iterated matrix games is to model each agent as an RL agent that independently attempts to maximize its expected total discounted reward. Several approaches to model agents in this way use policy gradient-based methods~\cite{sutton2000policy, williams1992simple}. Policy gradient methods update an agent's policy, parameterized by $\theta^{a}$, by performing gradient ascent on the expected total discounted reward $\E [R_{0}^{a}]$. Formally, let $\theta^{a}$ denote the parameterized version of an agent's policy $\pi^{a}$ and $V_{\theta^{1},\theta^{2}}^{a}$ denote the total expected discounted reward for agent $a$.  Here, $V^{a}$ is a function of the policy parameters $(\theta^{1}, \theta^{2})$ of both agents. In the $i^{th}$ iteration of the game, each agent updates $\theta_{i}^{a}$ to $\theta_{i+1}^{a}$, such that it maximizes it's total expected discounted reward. $\theta_{i+1}^{a}$ is computed using Eq.~\ref{eq:theta_argmax}. For agents trained using reinforcement learning, the gradient ascent rule to update $\theta_{i+1}^{1}$ is given by Eq.~\ref{eq:naive_learner_update_rule}.
\begin{equation}
    \small
        \theta_{i+1}^{1} = \argmax_{\theta^{1}} V^{1}(\theta^{1}, \theta_{i}^{2}) 
        ~~~\text{and}~~~
        \theta_{i+1}^{2} = \argmax_{\theta^{2}} V^{2}(\theta_{i}^{1}, \theta^{2})
    \label{eq:theta_argmax}%
\end{equation}\vspace{-5pt}%
\begin{equation}%
    \small
        f_{nl}^{1} = \nabla_{\theta^{i}_{1}} V^{1} (\theta_{i}^{1}, \theta_{i}^{2}) \cdot \delta ~~~\text{and}~~~
        \theta_{i+1}^{1} = \theta_{i}^{1} + f_{nl}^{1} (\theta_{i}^{1}, \theta_{i}^{2}) 
    \label{eq:naive_learner_update_rule}%
\end{equation}
where $\delta$ is the step size of the updates.
In the Iterated Prisoner's Dilemma (IPD) game, agents trained with the policy gradient update method converge to a sub-optimal mutual defection equilibrium (Figure~\ref{fig:results_IPD}, \citet{Peysakhovich1707.01068}). This sub-optimal equilibrium attained by Selfish Learners motivates us to explore alternative methods that could lead to a desirable cooperative equilibrium. We denote the agent trained using policy gradient updates as a Selfish Learner ($SL$).

\subsection{Learning Policies in Iterated Matrix Games: The Status-Quo Aware Learner ($SQLoss$)}
\label{sec:approach:SQLoss}
\subsubsection{$SQLoss$: Motivation and Theory}
\label{sec:approach:SQLoss:intuition}
\begin{figure}[hbt!]
    \centering
    \includegraphics[width=0.47\linewidth]{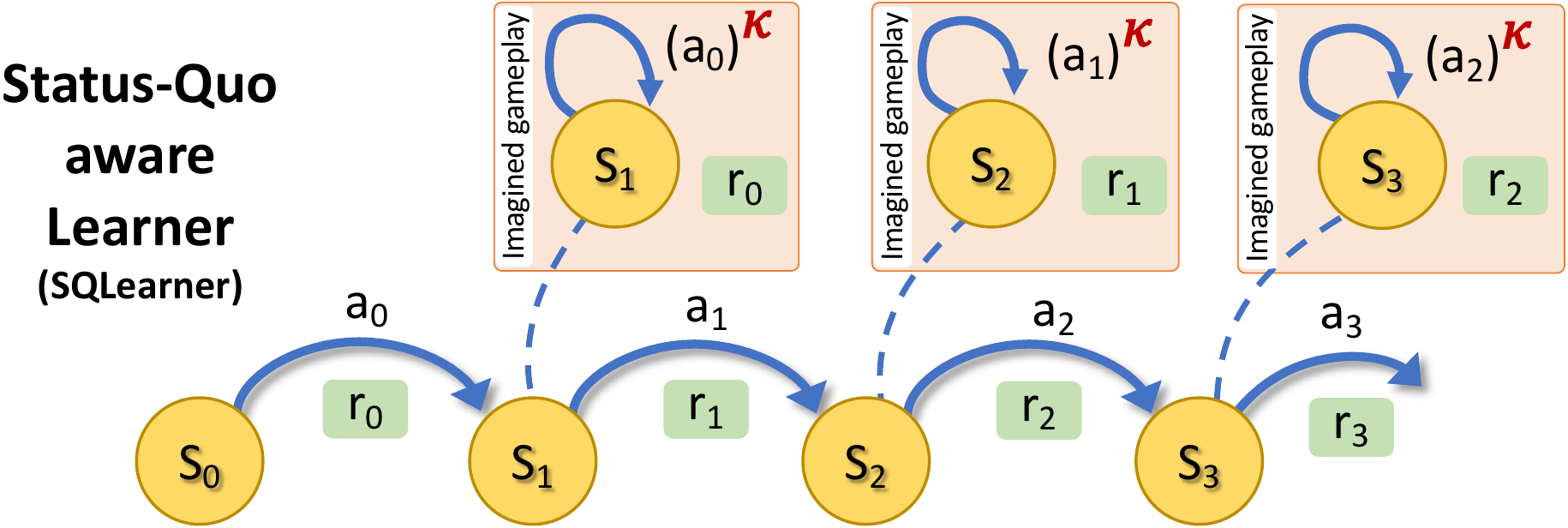}
    \caption{Intuition behind \textit{Status-Quo}-aware learner. At each step, the $SQLoss$ encourages an agent to imagine the consequences of sticking to the status-quo by imagining an episode where the status-quo is repeated for $\kappa$ steps. Section~\ref{sec:approach:SQLoss} describes $SQLoss$ in more detail.}
    \label{fig:rl_imaginative_learning}
\end{figure}

The status-quo bias instills in humans a preference for the current state provided the state is not harmful to them. 
Inspired by this idea, we introduce a status-quo loss ($SQLoss$) for each agent, derived from the idea of imaginary game-play (Figure~\ref{fig:rl_imaginative_learning}). 
The $SQLoss$ encourages an agent to imagine a future episode where the status-quo (current situation) is repeated for several steps. 
If an agent has been exploited in the previous iteration of the game (state \textit{DC}), then $SQLoss$ will encourage the agent to imagine a continued risk of exploitation and subsequently switch to defection and move to state \textit{DD}.  
Hence, for the exploiting agent, the short-term gain from exploitation (\textit{DC}) is overcome by the long-term loss from mutual defection (\textit{DD}). 
Conversely, if both agents cooperated in the previous iteration of the game (state \textit{CC}), then $SQLoss$ will encourage the agent to imagine a continued gain from mutual cooperation and subsequently stick to state \textit{CC}.
Since state \textit{CC} is more beneficial to both agents than \textit{DD}, and exploitation (\textit{DC} and \textit{CD}) also leads to \textit{DD}, agents move away from defection and converge to mutual cooperation (\textit{CC}). 
Appendix \ref{appendix:sqloss_theory} describes in detail how cooperation emerges between $SQLoss$ agents.

\subsubsection{$SQLoss$: Formulation}
\label{sec:approach:SQLoss:formulation}
We describe below the formulation of SQLoss with respect to agent 1. The formulation for agent 2 is identical to that of agent 1. Let $\uptau_a = (s_{0}, u_{0}^{1}, u_{0}^{2}, r_{0}^{1}, \cdots s_{T}, u_{T}^{1}, u_{T}^{2}, r_{T}^{1})$ denote the collection of an agent's experiences after $T$ time steps. Let $R_{t}^{1} (\uptau_1) = \sum_{l=t}^{T} \gamma^{l-t} r_{l}^{1}$ denote the discounted future return for agent $1$ starting at $s_t$ in actual game-play. Let $\hat \uptau_{1}$ denote the collection of an agent's \textbf{imagined} experiences. For a state $s_{t}$, where $t \in [0, T]$, an agent imagines an episode by starting at $s_{t}$ and repeating $u_{t-1}^{1}, u_{t-1}^{2}$ for $\kappa_{t}$ steps. This is equivalent to imagining a $\kappa_{t}$ step repetition of already played actions. We sample $\kappa_{t}$ from a Discrete Uniform distribution $\mathbb{U} \{1,z\}$ where $z$ is a hyper-parameter $\geq 1$. To simplify notation, let $\phi_t(s_t, \kappa_{t})$ denote the ordered set of state, actions, and rewards starting at time $t$ and repeated $\kappa_{t}$ times for imagined game-play. Let $\hat R_{t}^{1} (\hat \uptau_1)$ denote the discounted future return starting at $s_t$ in imagined status-quo game-play.
{
    \begin{equation}%
        \small
        \phi_t(s_t,\kappa_{t}) = \big[ (s_{t}, u_{t-1}^{1}, u_{t-1}^{2}, r_{t-1}^{1})_0,
        ~\cdots (s_{t}, u_{t-1}^{1}, u_{t-1}^{2}, r_{t-1}^{1})_1, (s_{t}, u_{t-1}^{1}, u_{t-1}^{2}, r_{t-1}^{1})_{\kappa_{t}-1} \big]
    \end{equation}%
}%
{%
    \begin{equation}%
        \small
        \hat \uptau_{1} = \big( \phi_t(s_t,\kappa_{t}), (s_{t+1}, u_{t+1}^{1}, u_{t+1}^{2}, r_{t+1}^{1})_{\kappa_{t} + 1}, ~\cdots~, (s_{T}, u_{T}^{1}, u_{T}^{2}, r_{T}^{1})_{T+\kappa_{t}-t} \big)
    \end{equation}%
}%
\begin{equation}%
    \small
    \hat R_{t}^{1} (\hat \uptau_1) = \Big( \frac{1 - \gamma^{\kappa}}{1 - \gamma} \Big) r_{t-1}^{1} + \gamma^{\kappa} R_{t}^{1} (\uptau_1) = \Big( \frac{1 - \gamma^{\kappa}}{1 - \gamma} \Big) r_{t-1}^{1} + \gamma^{\kappa} \sum_{l=t}^{T} \gamma^{l-t} r_{l}^{1}
    \label{eq:returns_hat}
\end{equation}%
$V_{\theta^{1},\theta^{2}}^{1}$ and $\hat V_{\theta^{1},\theta^{2}}^{1}$ are approximated by $\E [R_{0}^{1} (\uptau_1)]$ and $\E[\hat R_{0}^{1} (\hat \uptau_1)]$ respectively. These $V$ values are the expected rewards conditioned on both agents' policies ($\pi^{1}$, $\pi^{2}$). For agent 1, the regular gradients and the Status-Quo gradient-like correction term, $\nabla_{\theta^{1}} \E [R_{0}^{1}(\uptau_1)]$ and $f_{sqcor}^{1}$, can be derived from the policy gradient formulation as
\begin{equation}
    \small
    \begin{split}
        \nabla_{\theta^{1}} \E [R_{0}^{1}(\uptau_1)] = \E  [R_{0}^{1}(\uptau_1)\nabla_{\theta^{1}} \log\pi^{1}(\uptau_1)]
        & = \E {\Big[} \sum_{t=1}^{T} \nabla_{\theta^{1}}  \log\pi^{1} (u_{t}^{1}|s_{t}) \gamma^{t} \big( R_{t}^{1}(\uptau_1) - b(s_{t}) \big) \Big]
    \end{split}
    \label{eq:pg_original}
\end{equation}%
\vspace{-7pt}
\begin{equation}%
    \small
    \begin{split}%
        f_{sqcor}^{1}
%
        & = \E {\Bigg[} \sum_{t=1}^{T} \nabla_{\theta^{1}} \log\pi^{1} (u_{t-1}^{1}|s_{t}) \times
        \gamma^{t} \times \Bigg( \Big( \frac{1 - \gamma^{\kappa}}{1 - \gamma} \Big) r_{t-1}^{1} + \gamma^{\kappa} \sum_{l=t}^{T} \gamma^{l - t} r_{l}^{1} - b(s_{t}) \Bigg) \Bigg] \\
        &= \E {\Big[} \sum_{t=1}^{T} \nabla_{\theta^{1}} \log\pi^{1} (u_{t-1}^{1}|s_{t}) \times \gamma^{t} \times \big( \hat R_{t}^{1}(\hat \uptau_1) - b(s_{t}) \big) \Big]
    \end{split}
    \label{eq:status_quo_pg}
\end{equation}
Here, $f_{sqcor}^{1}$ in Eq.~\ref{eq:status_quo_pg} is a biased term that has policy gradient-like structure and incorporates the prior action taken through the $u_{t-1}|s_t$ term.
It should be noted that the $f_{sqcor}^{1}$ is not a gradient of the expected return of the imagined trajectory in Eq.~\ref{eq:returns_hat}, but the form of this expression is arrived at considering the policy gradient expression. 
Furthermore, $b(s_t)$ is a baseline-like term added for variance reduction in the implementation. 

The update rule $f_{sql,pg}$ for the policy gradient-based Status-Quo Learner (SQL-PG) is defined by,
\begin{equation}
    f_{sql,pg}^{1} = \big( \alpha \cdot \nabla_{\theta^{1}} \E [R_{0}^{1}(\uptau_1)] + \beta \cdot f_{sqcor}^{1} \big) \cdot \delta
\end{equation}
where $\alpha$, $\beta$ are the loss scaling factor for REINFORCE and imaginative game-play respectively.
\begin{table}
    \centering
    \small
    \begin{subtable}[t]{0.3\linewidth}
        \centering
        \small
        \renewcommand{\arraystretch}{1.1}
        \begin{tabular}{c|c|c}
             & \textit{C} & \textit{D} \\ \hline
             \textit{C} & (-1, -1) & (-3, 0) \\ \hline
             \textit{D} & (0, -3) & (-2, -2) \\ \hline
        \end{tabular}
        \caption{Prisoners' Dilemma (PD)}
        \label{tab:payoff_matrix_ipd}
    \end{subtable}%
    \begin{subtable}[t]{0.3\linewidth}
        \centering
        \small
        \renewcommand{\arraystretch}{1.1}
        \begin{tabular}{c|c|c}
             & \textit{H} & \textit{T} \\ \hline
             \textit{H} & (+1, -1) & (-1, +1) \\ \hline
             \textit{T} & (-1, +1) & (+1, -1) \\ \hline
        \end{tabular}
        \caption{Matching Pennies (MP)}
        \label{tab:payoff_matrix_imp}
    \end{subtable}%
    \begin{subtable}[t]{0.3\linewidth}
        \centering
        \small
        \renewcommand{\arraystretch}{1.1}
        \begin{tabular}{c|c|c}
             & \textit{C} & \textit{D} \\ \hline
             \textit{C} & (0, 0) & (-4, -1) \\ \hline
             \textit{D} & (-1, -4) & (-3, -3) \\ \hline
        \end{tabular}
        \caption{Stag Hunt (SH)}
        \label{tab:payoff_matrix_sh}
    \end{subtable}%
    \caption{\label{tab:payoff_matrix}%
    Payoff matrices for the different games used in our experiments. $(X,Y)$ in a cell represents a reward of $X$ to the row and $Y$ to the column player. \textit{C}, \textit{D}, \textit{H}, and \textit{T} denote the actions for the row and column players. In the iterated versions of these games, agents play against each other over several iterations. In each iteration, an agent takes an action and receives a reward based on the actions of both agents. Each matrix represents a different kind of social dilemma.}%
    \vspace{-13pt}
\end{table}%

\subsection{Learning policies in Dynamic Non-Matrix Games using $SQLoss$ and \ipdistill}
\label{sec:GameDistill}
The previous section focused on learning optimal policies in iterated matrix games that represent sequential social dilemmas. 
In such games, an agent can take one of a discrete set of policies at each step. 
For instance, in IPD, an agent can either cooperate or defect at each step. 
However, in social dilemmas such as the Coin Game (Appendix~\ref{appendix:game-details}), cooperation and defection policies are composed of a sequence of state-dependent actions.
$SQLoss$, proposed above, works for matrix games but is not directly applicable to games with visual input to yield mutual cooperation.
To apply the Status-Quo policy gradient to these games, we present \ipdistill, a self-supervised algorithm that reduces a dynamic infinitely iterated game with visual input to a matrix game.
\ipdistill takes as input game-play episodes between agents with random policies and learns oracles (or policies) that lead to distinct outcomes. 
\ipdistill (Figure~\ref{fig:high_level_approach}) works as follows. 
\begin{enumerate}%
    \item We initialize agents with random weights and play them against each other in the game. In these \textbf{random game-play} episodes, whenever an agent receives a reward, we store the sequence of states along with the rewards for both agents. %
    \item This collection of state sequences is used to train the \ipdistill network, which is a \textbf{self-supervised trajectory encoder}. It takes as input a sequence of states and predicts the rewards of both agents during training.%
    \item We now \textbf{cluster the embeddings} extracted from the penultimate layer of the trained \ipdistill network using Agglomerative Clustering~\cite{friedman2001elements}.
    Each embedding is a finite-dimensional representation of the corresponding state sequence.
    Each cluster represents a collection of state sequences or transitions, that lead to a consistent outcome (w.r.t rewards). 
    For CoinGame, when we use the number of clusters as two, we observe that one cluster consists of transitions that represent cooperative behavior (cooperation cluster) while the other contains transitions that lead to defection (defection cluster). 
    \item Using the state sequences in each cluster, we \textbf{train an oracle} to predict the next action given the current state. 
    For the Coin Game, the oracle trained on state sequences from the cooperation cluster predicts the cooperative action for a given state.
    Similarly, the oracle trained on the defection cluster predicts the defection action for a given state.
    Each agent uses \ipdistill independently to extract a cooperation and a defection oracle.
    Figure~\ref{fig:oracle_predictions} (Appendix~\ref{appendix:oracle:illustration}) illustrates the cooperation and defection oracles extracted by the Red agent.%
\end{enumerate}

During game-play, an agent can consult either oracle at each step. 
In CoinGame, this is equivalent to either cooperating (consulting the cooperation oracle) or defecting (consulting the defection oracle). 
In this way, an agent reduces a dynamic game to its matrix equivalent using \ipdistill. 
We then utilize the Status-Quo policy gradient to learn optimal policies in the reduced matrix game.
For the Coin Game, this leads to agents who cooperate by only picking coins of their color (Figure~\ref{fig:coingame_results}).
It is important to note that for games like CoinGame, we could have learned cooperation and defection oracles by training agents using the sum-of-rewards for both agents and individual reward, respectively~\cite{Peysakhovich1707.01068}. 
However, \ipdistill learns distinct policies without using hand-crafted reward functions.
\ipdistill is applicable only in games where cooperation and defection policies are clearly defined and lead to distinct payoffs (rewards) for both players.

Algorithm~\ref{algorithm:pseudo_code_gamedistill:algorithm1} in Appendix provides the pseudo-code for \ipdistill with additional details in Appendix~\ref{appendix:gamedistill:pseudo-code}.
Appendix~\ref{appendix:gamedistill:architecture-details} provides additional details about the architecture and the different components of \ipdistill. 
Further, we have verified \ipdistill empirically on the Coin Game and the Stag Hunt (results in Appendix~\ref{appendix:results:visual-staghunt-with-gamedistill-and-sqloss}). 
Additional clustering visualizations for the trajectories and the experimental plots are provided in Appendix~\ref{appendix:gamedistill-clustering} and~\ref{appendix:oracle:illustration}.


\section{Experimental Setup}
\label{sec:experimental_setup}
In order to compare our results to previous work, we use the Normalized Discounted Reward or $NDR = (1-\gamma) \sum_{t=0}^{T} \gamma^{t} r_{t}$. 
A higher NDR implies that an agent obtains a higher reward in the environment. 
We compare our approach (Status-Quo Aware Learner or $SQLearner$) to Learning with Opponent-Learning Awareness (Lola-PG)~\cite{foerster2018learning} and the Selfish Learner (SL) agents.
For all experiments, we perform $20$ runs and report average $NDR$, along with variance across runs. 
The bold line in all the figures is the mean, and the shaded region is the one standard deviation region around the mean.

\subsection{Iterated Matrix Game Social Dilemmas}
For our experiments with social dilemma matrix games, we use the Iterated Prisoners Dilemma (IPD)~\cite{luce1989games}, Iterated Matching Pennies (IMP)~\cite{lee1967application}, and the Iterated Stag Hunt (ISH)~\cite{fang2002adaptive}. 
Each matrix game in Table~\ref{tab:payoff_matrix} represents a different dilemma.  
In the Prisoner's Dilemma, the rational policy for each agent is to defect, regardless of the other agent's policy.
However, when each agent plays rationally, each is worse off. 
In Matching Pennies, if an agent plays predictably, it is prone to exploitation by the other agent. 
Therefore, the optimal policy is to randomize between \textit{H} and \textit{T}, obtaining an average NDR of $0$. 
The Stag Hunt game represents a coordination dilemma. 
In the game, given that the other agent will cooperate, an agent's optimal action is to cooperate as well. 
However, each agent has an attractive alternative at each step, that of defecting and obtaining a guaranteed reward of $-1$. 
Therefore, the promise of a safer alternative and the fear that the other agent might select the safer choice could drive an agent to select the safer alternative, thereby sacrificing the higher reward of mutual cooperation. 

In iterated matrix games, at each iteration (iter), agents take an action according to a policy and receive the rewards in Table~\ref{tab:payoff_matrix}. 
To simulate an infinitely iterated game, we let agents play 200 iters of game against each other, and do not provide an agent with any information about the number of remaining iters. 
In an iter, state for an agent is the actions played by both agents in the previous iter.

\subsection{Iterated Dynamic Game Social Dilemmas}
For our experiments on a social dilemma with extended actions, we use the Coin Game (Figure~\ref{fig:coin_game})~\cite{foerster2018learning} and the non-matrix variant of the Stag Hunt (Figure~\ref{fig:stag_hunt_game}). 
We provide details of the games in Appendix~\ref{appendix:game-details} due to space considerations.
\section{Results}
\label{sec:results}

\subsection{Learning optimal policies in Iterated Matrix Dilemmas}
\label{sec:results:iterated-matrix-games:sqloss}

\begin{figure}[!htb]
    \begin{minipage}{0.45\textwidth}
        \centering
        \includegraphics[width=0.62\linewidth]{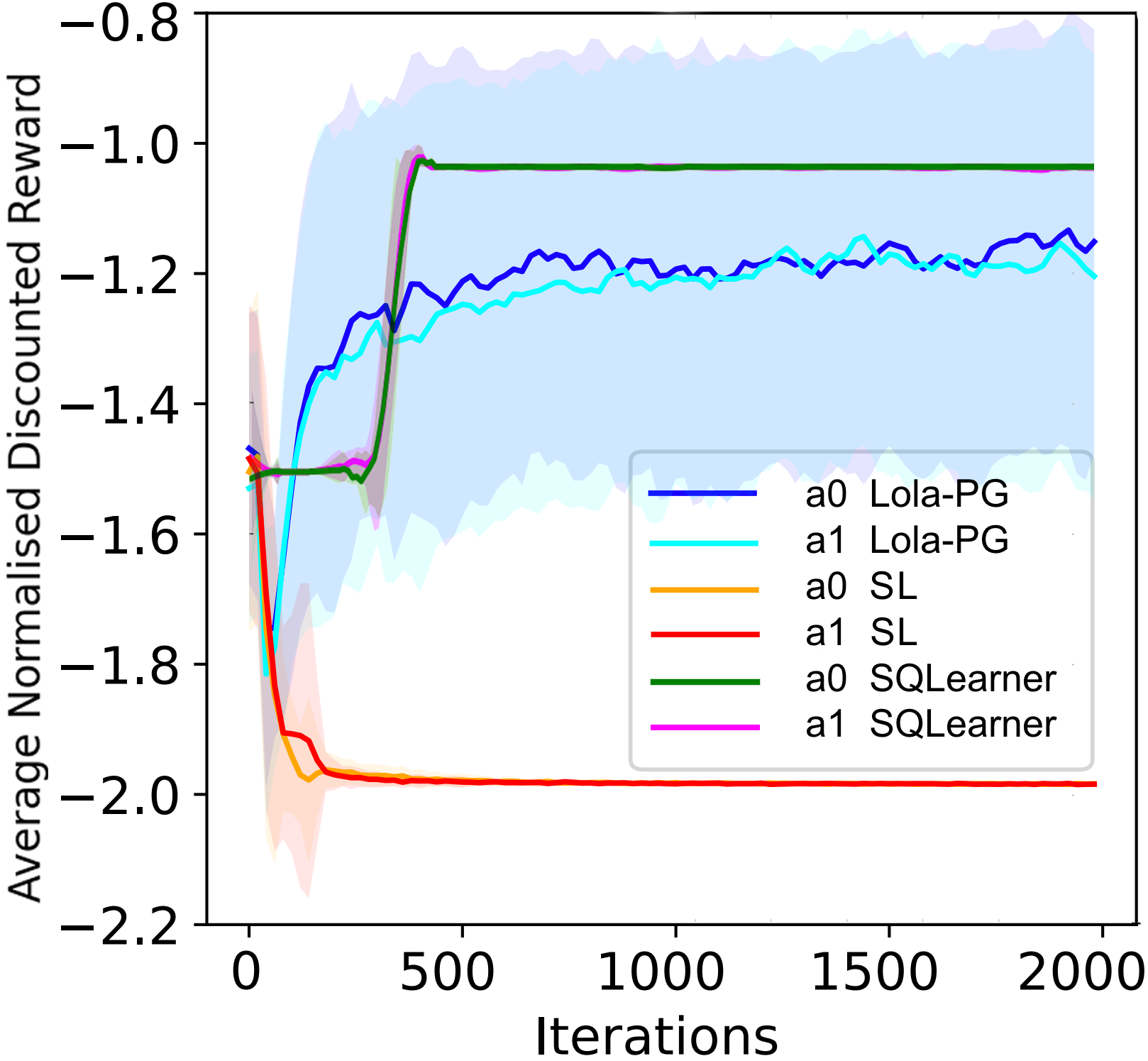}
        \caption{\label{fig:results_IPD}%
        Average NDR values for different learners in IPD. $SQLearner$ agents obtain a near-optimal NDR value $(-1)$ for this game.}
    \end{minipage}\hfill
    \begin{minipage}{0.53\textwidth}
        \centering
        \includegraphics[width=0.92\linewidth]{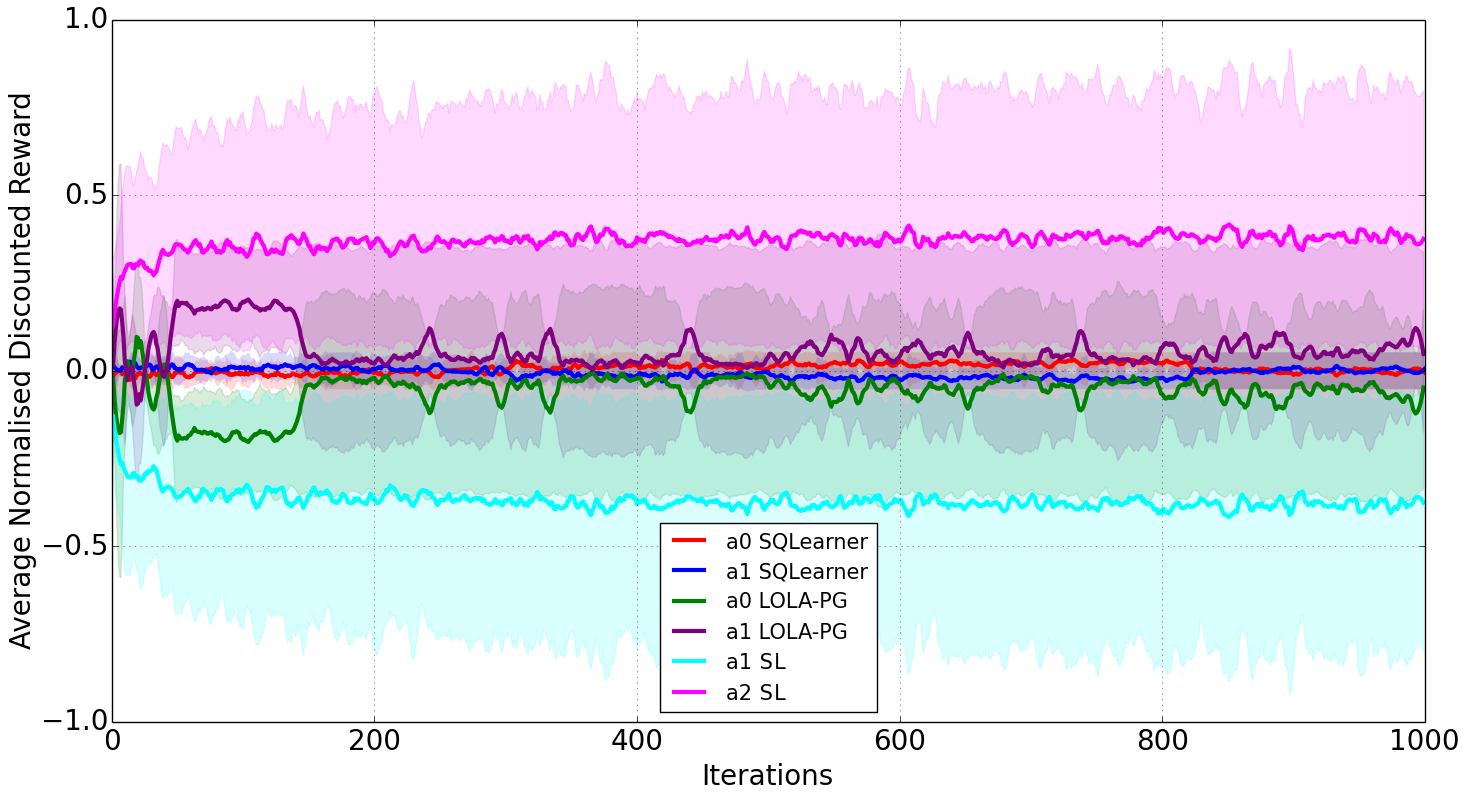}
        \caption{\label{fig:results_IMP}%
        Average NDR values for different learners in IMP. $SQLearner$ agents avoid exploitation by randomising between \textit{H} and \textit{T} to obtain a near-optimal NDR value (0).}
    \end{minipage}\hfill%
\end{figure}

\textbf{Iterated Prisoner's Dilemma (IPD):}
We train different learners to play the IPD game. 
Figure~\ref{fig:results_IPD} shows the results. 
For all learners, agents initially defect and move towards an NDR of $-2.0$. 
This initial bias towards defection is expected, since, for agents trained with random game-play episodes, the benefits of exploitation outweigh the costs of mutual defection. 
For Selfish Learner (SL) agents, the bias intensifies, and the agents converge to mutually harmful selfish behavior (NDR of $-2.0$). 
Lola-PG agents learn to predict each other's behavior and realize that defection is more likely to lead to mutual harm. 
They subsequently move towards cooperation, but occasionally defect (NDR of $-1.2$). 
In contrast, $SQLearner$ agents quickly realize the costs of defection, indicated by the small initial dip in the NDR curves.
They subsequently move towards close to 100\% cooperation, with an NDR of $-1.0$.
Finally, it is important to note that $SQLearner$ agents have close to zero variance, unlike other methods where the variance in NDR across runs is significant.

\textbf{Iterated Matching Pennies (IMP):}
We train different learners to play the IMP game. 
Figure~\ref{fig:results_IMP} shows the results. 
$SQLearner$ agents learn to play optimally and obtain an NDR close to $0$. 
Interestingly, Selfish Learner (SL) and Lola-PG agents converge to an exploiter-exploited equilibrium where one agent consistently exploits the other agent. 
This asymmetric exploitation equilibrium is more pronounced for SL agents than for Lola-PG agents. 
As before, we observe that $SQLearner$ agents have close to zero variance across runs, unlike other methods where the variance in NDR is significant.

\textbf{Iterated Chicken Game (ICG):} Appendix~\ref{appendix:results:chicken} shows additional results for the ICG game.
\begin{figure}[!htb]
    \begin{minipage}{0.4\textwidth}
        \centering
        \includegraphics[width=0.65\linewidth]{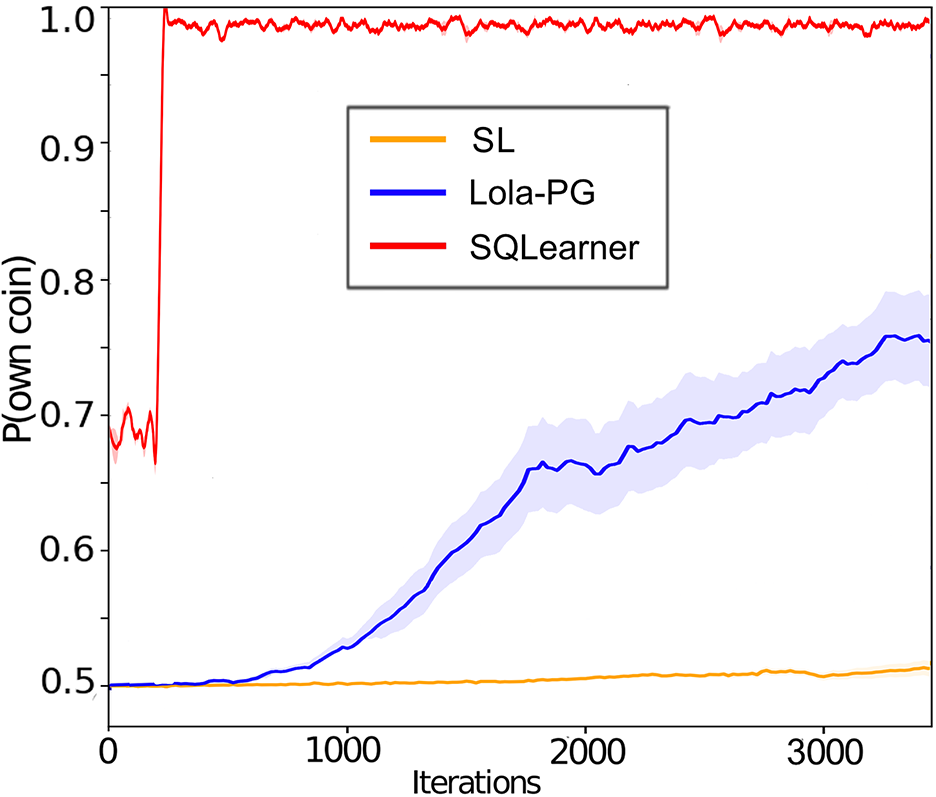}
        \caption{\label{fig:coingame_results}%
        Probability that an agent will pick a coin of its color in CoinGame.
        $SQLearner$ agents achieve a cooperation rate close to 1.0 while Lola-PG agents achieve rate close to 0.8.}
    \end{minipage}\hfill%
    \begin{minipage}{0.55\textwidth}
        \centering
        \includegraphics[width=0.85\linewidth]{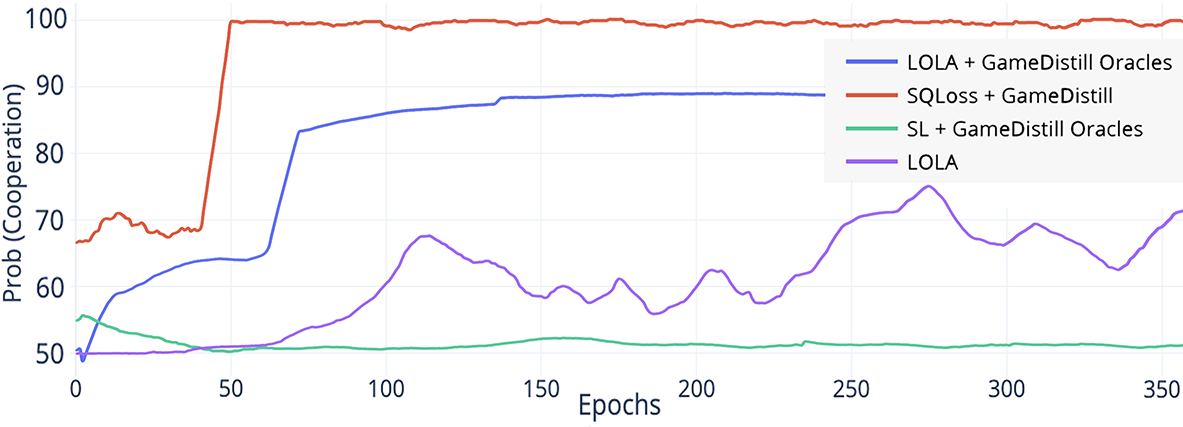}
        \caption{\label{fig:gamedistill_oracles_with_lola_and_others}%
        Impact of using pretrained \ipdistill oracles in CoinGame. 
        $SQLearner$ agents (w/ \ipdistill oracles) achieve close to 100\% cooperation. 
        Lola-PG agents eventually (after roughly 4000 epochs) achieve close to 80\% cooperation.
        Interestingly, Lola-PG agents trained w/ \ipdistill oracles converge to higher cooperation rate of 85\% in 300 epochs.}%
    \end{minipage}\hfill%
\end{figure}%

\subsection{Learning Optimal Policies in Iterated Dynamic Dilemmas}
\label{sec:results:visual-input-gamedistll-and-sqloss}

\textbf{GameDistill:} To evaluate the clustering step in \ipdistill, we make two t-SNE~\cite{maaten2008visualizing} plots of the 100-dimensional feature vectors extracted from the penultimate layer of the trained \ipdistill network in Figure~\ref{fig:feature_vector_plots:all} of Appendix~\ref{appendix:gamedistill-clustering}. 
In the first plot, we color each point (or state sequence) by the rewards obtained by both agents in the format $r_1|r_2$.
In the second, we color each point by the cluster label output by the clustering technique.
\ipdistill correctly discovers two clusters, one for transitions that represent cooperation (Red cluster) and the other for transitions that represent defection (Blue cluster). 
We experiment with different values for feature vector dimensions and clustering techniques, and obtain similar results (see Appendix~\ref{appendix:experimental-setup:gamedistill}).
Results on Stag Hunt using \ipdistill are presented in Appendix~\ref{appendix:gamedistill-clustering} and~\ref{appendix:results:visual-staghunt-with-gamedistill-and-sqloss}.
To evaluate the trained oracles that represent cooperation and a defection policy, we modify the CoinGame environment to contain only a single agent (the Red agent). 
We then play two variations of the game.
In the first variant, the Red agent executes the actions as suggested by the first oracle. 
We observe that Red agent picks only $8.4\%$ of Blue coins, indicating a high cooperation rate, which represents a cooperation policy. 
In the second variant, Red agent executes the actions suggested by the second oracle. 
We observe that Red agent picks $99.4\%$ of Blue coins, indicating high defection rate, which represents a defection policy.

\textbf{SQLoss:}
During game-play, at each step, an agent follows either the action suggested by its cooperation oracle or defection oracle. 
We compare approaches using the degree of cooperation between agents, measured by the probability that an agent will pick the coin of its color~\cite{foerster2018learning}.
Figure~\ref{fig:coingame_results} shows the results. 
The probability that an $SQLearner$ agent will pick the coin of its color is close to $1.0$. 
This high probability indicates that the other $SQLearner$ agent is cooperating with this agent and only picking coins of its color. 
In contrast, the probability that a Lola-PG agent will pick a coin of its color is close to $0.8$, indicating higher defection rates. 
As expected, the probability of an agent picking its own coin is the smallest for Selfish Learner. 

To evaluate the impact of \ipdistill, we compare $SQLearner$ agents to Lola-PG agents trained using \ipdistill oracles.
Figure~\ref{fig:gamedistill_oracles_with_lola_and_others} shows the results.
Interestingly, Lola-PG agents trained with \ipdistill oracles converge to a higher cooperation rate ($0.88$) than Lola-PG agents trained directly on visual input.

\section{$SQLoss$ for social dilemma matrix games}
\label{sec:extending_sqloss_with_liebo_conditions}
For a given matrix game, $SQLoss$ will work on an equivalent version of the game in which all rewards have been transformed to non-positive values. For the matrix games described in our paper, we have used their variants with negative rewards to remain consistent with the LOLA paper. For a general matrix game, we can subtract the maximum reward (or any number larger than it) from each reward value to make rewards negative and then use $SQLoss$. We consider the social dilemma class of matrix games from \citet{Leibo:2017}, $%
\scriptsize
\begin{pmatrix}
    & C & D \\ 
    C & R,R & S,T \\
    D & T,S & P,P \\
\end{pmatrix}
$,%
where the first row corresponds to cooperation for the first player and the first column corresponds to cooperation for the second player. \citet{Leibo:2017} define the following rules that describe different categories of social dilemmas: \textbf{(i)} If $R > P$, then mutual cooperation is preferred to mutual defection. \textbf{(ii)} If $R > S$, then mutual cooperation is preferred to being exploited by a defector. \textbf{(iii)} If $2R > T + S$, this ensures that mutual cooperation is preferred to an equal probability of unilateral cooperation and defection. \textbf{(iv)} Either greed ($T > R$: Exploit a cooperator preferred over mutual cooperation) or fear ($P > S$: Mutual defection preferred over being exploited) should hold.
These rules have been reproduced from \citet{Leibo:2017},%
\begin{enumerate}%
    \item When ``greed'' ($T > R$) as well as ``fear'' ($P > S$) conditions hold, we have $T > R > P > S$. The Iterated Prisoner's Dilemma (IPD) is an example of this game. If we subtract $T$ from each reward, we get the matrix game, $%
        \scriptsize
        \begin{pmatrix}
            & C & D \\
            C & R-T,R-T & S-T,0 \\
            D & 0,S & P-T,P-T \\
        \end{pmatrix}
    $,%
    where all entries are non-positive and therefore Lemma 3 and Lemma 4 hold as before.
    \item When ``greed'' holds but not ``fear'' we have $T > R > S >= P$. The Chicken Game (CG) is an example of this game. If we subtract $T$ from each reward as before, we get the equivalent matrix game with non-positive entries and Lemmas 3 and 4 hold.
    \item When ``fear'' holds but not ``greed'' we have $R >= T > P > S$. The Iterated Stag Hunt is an example of this game. If we subtract $R$ from each reward, we get the equivalent matrix game with non-positive entries then Lemmas~\ref{lemma:lemma3} and~\ref{lemma:lemma4} (Appendix~\ref{appendix:sqloss_theory}) hold.
\end{enumerate}
In our experiments, we have considered games from each of the classes mentioned above, and the use of $SQLoss$ leads to cooperation in all these examples.

\section{Games with more than 2 players}
\label{sec:beyond_2_players}
Our formulation of $SQLoss$ has the distinct advantage of being fully ego-centric, i.e., the learning agent does not require any information regarding its opponents. This feature enables a straightforward extension of $SQLoss$ beyond the two agent setting, without any change in each agent's learning algorithm. In order to test this extension of $SQLoss$ beyond 2-players, we consider as an example, the problem described in the popular Braess' paradox\cite{Braess1968}, which is a well-known extension of the Prisoner's Dilemma problem to more than 2 agents. We performed additional experiment in this game with 4 agents. We simulated the result when all agents are selfish learners (the $SL$ agent(s)) and also when all agents use $SQLoss$ (the $SQLearner$(s)). We observe that when using selfish learners, all agents converge to Defection and when using $SQLoss$, all agents converge to Cooperation. Detailed explanation of the game, experimental setup and results are shown in Appendix~\ref{appendix:beyond_2_players}. 
\section{Conclusion}
We presented a status-quo policy gradient that encourages an agent to imagine the consequences of sticking to the status quo. 
We demonstrated how $SQLoss$ outperforms LOLA on standard benchmark matrix games. 
To work with dynamic games, we further proposed \ipdistill, an algorithm that reduces a dynamic game with visual input to a matrix game.
We combined \ipdistill and $SQLoss$ to demonstrate how agents learn optimal policies in dynamic social dilemmas with visual observations. 
We empirically demonstrated that $SQLoss$ obtains near-optimal rewards in various social-dilemma games such as IPD, IMP, Chicken, Stag-Hunt and Coin Game (both matrix and variant with visual observations).
\label{sec:conclusion}

\bibliographystyle{plainnat}
\bibliography{main-content/references}


\clearpage

\appendix
\section*{\textit{Appendix for}\\
Status-Quo Policy Gradient in Multi-agent Reinforcement Learning
}

\section{Description of Environments Used for Dynamic Social Dilemmas}
\label{appendix:game-details}
The three matrix games tested in the paper are canonical games that appear in the sequential social dilemma literature. Hence, we selected these to demonstrate the effectiveness of SQ policy gradient approach. We have not tested our approach beyond the social dilemma setting and hence limit our claims to the same.

\subsection{Coin Game}
\label{appendix:game-details:coin-game}
\begin{figure}[!htb]%
    \begin{minipage}{0.52\textwidth}
        \centering
        \includegraphics[width=\linewidth]{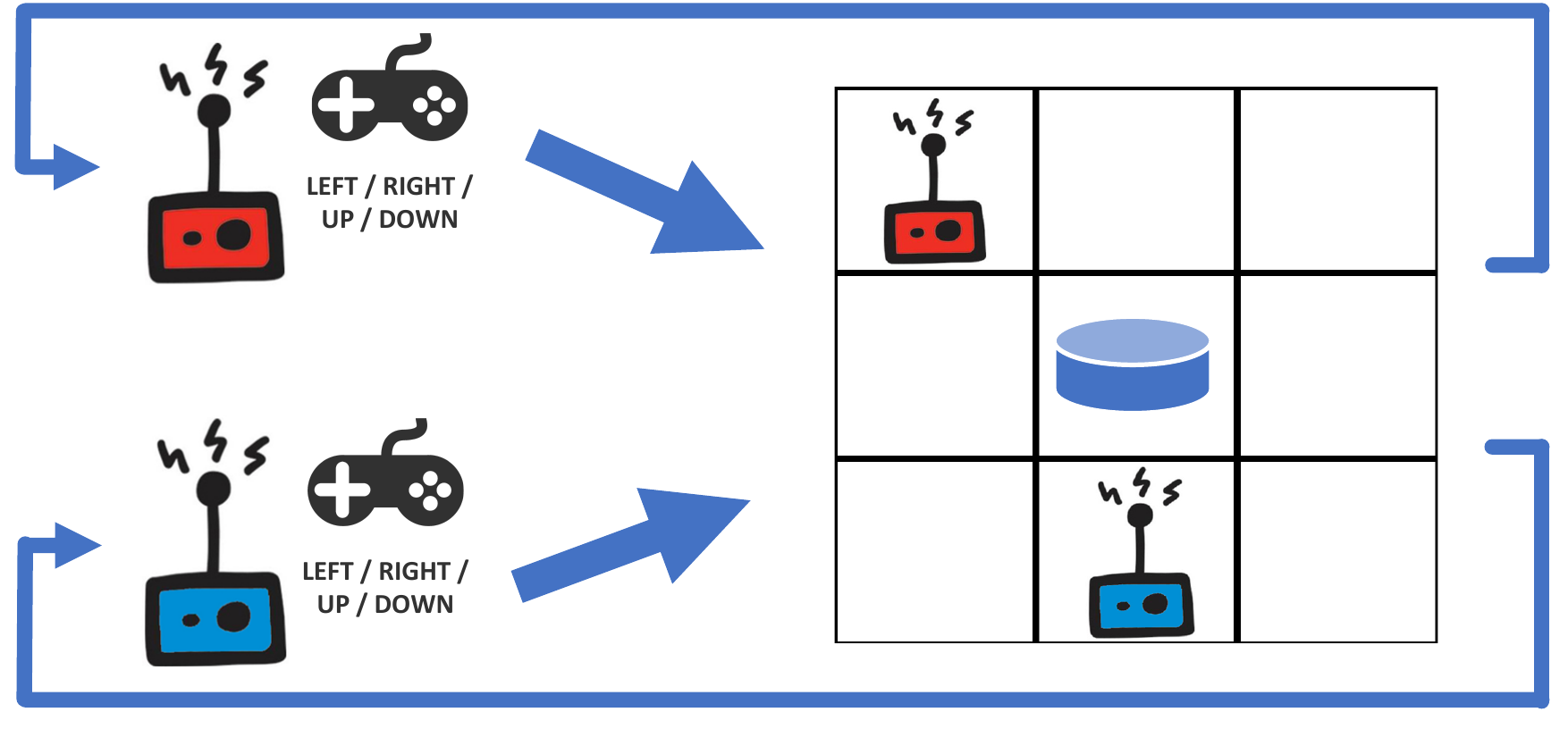}
        \caption{\label{fig:coin_game}%
        Illustration of two agents (Red and Blue) playing the dynamic game Coin Game}%
    \end{minipage}\hfill%
    \begin{minipage}{0.45\textwidth}
        \centering
        \includegraphics[width=\linewidth]{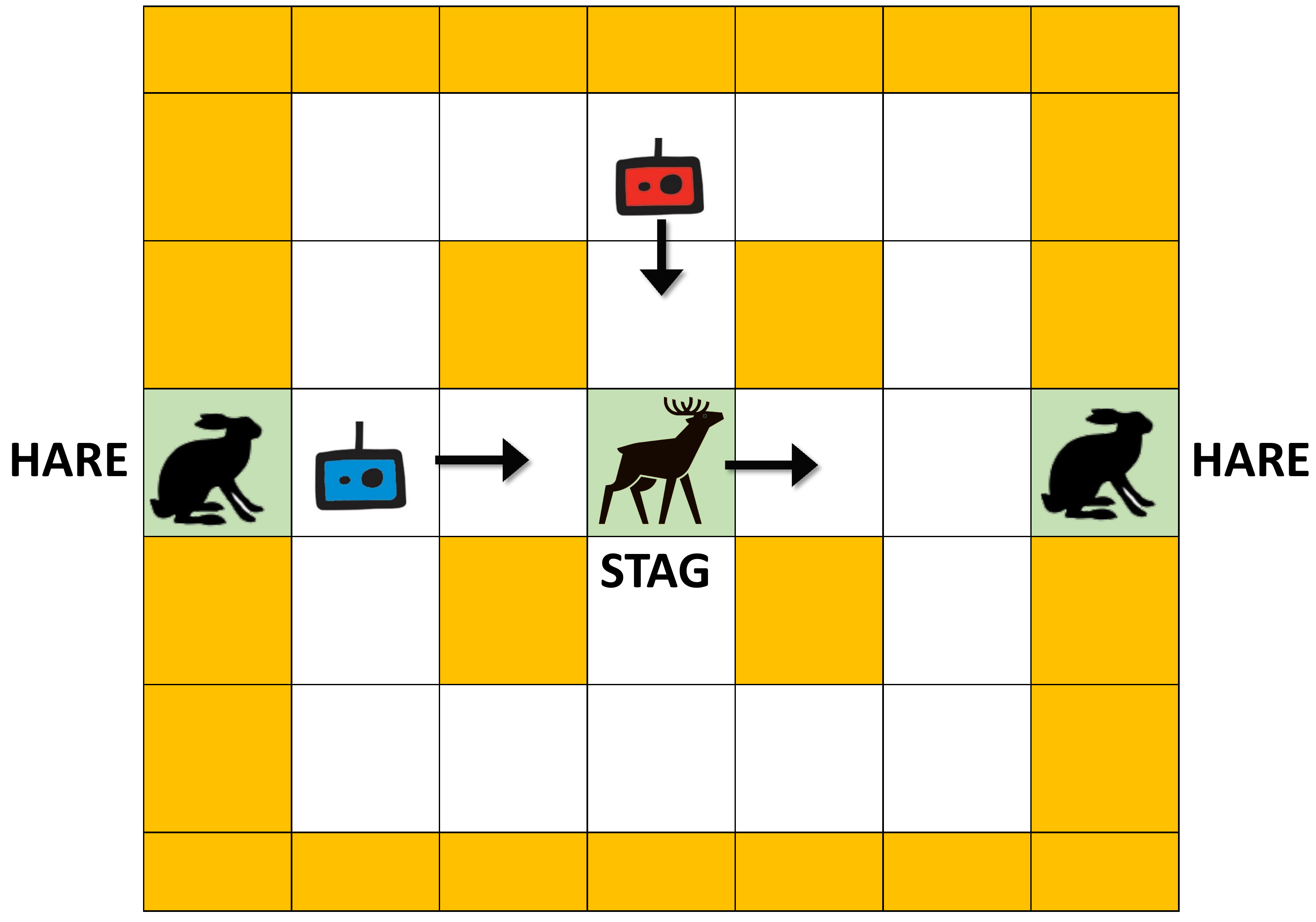}
        \caption{\label{fig:stag_hunt_game}%
        Illustration of two agents (Red and Blue) playing the dynamic game Stag-Hunt Game}%
    \end{minipage}\hfill%
\end{figure}

Figure~\ref{fig:coin_game} illustrates the agents playing the Coin Game. The agents, along with a Blue or Red coin, appear at random positions in a $3\times{}3$ grid. An agent observes the complete $3\times{}3$ grid as input and can move either left, right, up, or down. When an agent moves into a cell with a coin, it picks the coin, and a new instance of the game begins where the agent remains at their current positions, but a Red/Blue coin randomly appears in one of the empty cells. If the Red agent picks the Red coin, it gets a reward of $+1$, and the Blue agent gets no reward. If the Red agent picks the Blue coin, it gets a reward of $+1$, and the Blue agent gets a reward of $-2$. The Blue agent's reward structure is symmetric to that of the Red agent.
\subsection{Stag-Hunt}
\label{appendix:game-details:stag-hunt}
Figure~\ref{fig:stag_hunt_game} shows the illustration of two agents (Red and Blue) playing the visual Stag Hunt game. The STAG represents the maximum reward the agents can achieve with HARE in the center of the figure. An agent observes the full $7\times{}7$ grid as input and can freely move across the grid in only the empty cells, denoted by white (yellow cells denote walls that restrict the movement). Each agent can either pick the STAG individually to obtain a reward of $+4$, or coordinate with the other agent to capture the HARE and obtain a better reward of $+25$.

\section{\ipdistill: Oracles, Network Architecture and pseudo-code}
\label{appendix:experimental-setup:gamedistill}

\subsection{Oracles from \ipdistill}
\label{appendix:gamedistill:oracle-explanation}
The oracles are important in games with visual input. 
The agent uses these oracles to reduce the games to their matrix equivalents. 
While we call the two oracles learned from $GameDistill$ as Cooperation and Defection oracles, we do not need the notion of cooperation or defection, nor do we need to explicitly label these clusters as `cooperation' or `defection' to learn these oracles. 
These oracles are learned by clustering the outcomes of random play into two distinct clusters. 
In social dilemmas, these two distinct clusters represent cooperation and defection outcomes. 
Hence, we use the names `Cooperation' and `Defection' oracles for the oracles learned from these clusters.
It is important to mention that $SQLoss$ (without oracles) achieves high-degree of cooperation in matrix games. 
\begin{algorithm}[H]
    \caption{Pseduo-code for \ipdistill}
    \begin{algorithmic}[1]
        \STATE {\bfseries Input:} Game Environment $env$, Agents $agents$, Clustering Technique $AggClustering$
        \FOR{$agent$ {\bfseries in} $agents$}
            \STATE $t\_data=collect\_data(env,~agent)$ ~~~~~~~~~~~~~~~~~~~~~~~~~~~~~~// \textit{Collect trajectory data}\\
            \STATE $rewardPredNet = createNetwork()$
            \STATE $train\_network(rewardPredNet,~t\_data)$

            \STATE $feats = get\_features(rewardPredNet,~t\_data)$  ~~~~~~// \textit{Extract shared features \& cluster}
            \STATE $clus\_ids = AggClustering(n~$=$~2).fit(feats)$

            \STATE $oracle\_nets=[~]$  ~~~~~~~~~~~~~~~~~~~~~~~~~~~~~~~~~~~~~~~// \textit{Train oracles corresponding to each cluster}
            \FOR{$k$ {\bfseries in} range($2$)} 
              \STATE $index=np.where(clus\_ids~$==$~k)$
              \STATE $cluster\_data= t\_data[index]$
              \STATE $oracle\_nets[k]= create\_oracle\_net(env)$
              \STATE $train\_oracle(oracle\_nets[k],~cluster\_data)$
            \ENDFOR
        \ENDFOR
        \STATE  {\bfseries Output:} Trained oracle networks $oracle\_nets$
    \end{algorithmic}
    \label{algorithm:pseudo_code_gamedistill:algorithm1}
\end{algorithm}

In visual-input games with complex actions (such as up, down, left, right, eat-coin, etc.), it is not clear which action or sequence of actions constitute cooperation or defection.
In such games, the role of the cooperation oracle is to recommend, at each step, which action (out of up down, left, right, eat-coin, etc.) constitutes \textit{cooperative} behavior.
Similarly, the role of the defection oracle is to recommend, at each step, which action constitutes \textit{defection} behavior.
Algorithm~\ref{algorithm:pseudo_code_gamedistill:algorithm1} describes (at a high level) how agents train and use these oracles in the game-play life cycle.
Algorithm~\ref{algorithm:pseudo_code:train_oracles} (in Appendix~\ref{appendix:experimental-setup:gamedistill}) describes how the oracles are trained.

We also provide details about the oracle network architecture from in Appendix B.1 in the supplementary material.

\subsection{\ipdistill: Architecture Details}
\label{appendix:gamedistill:architecture-details}
\ipdistill consists of two components. 

\begin{figure*}[htb!]
    \centering
    \begin{subfigure}[t]{0.48\linewidth}
        \includegraphics[width=\linewidth]{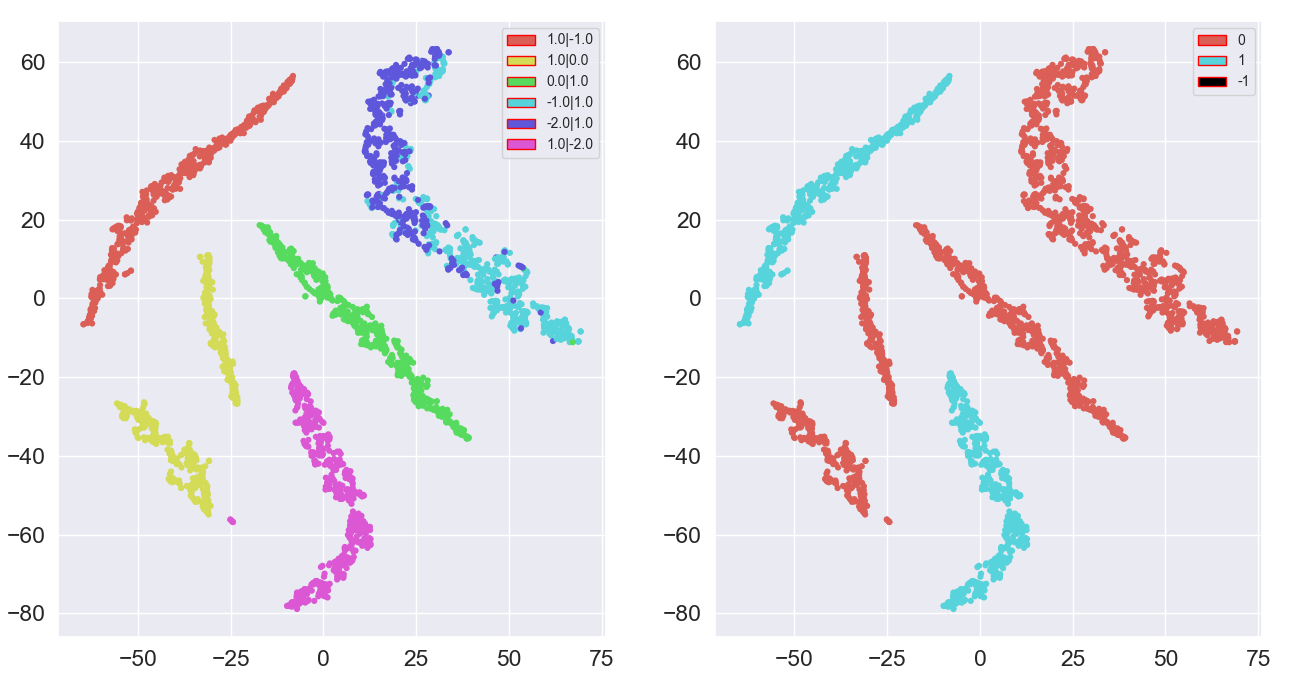}
        \caption{3-dimensional}
        \label{fig:feature_vector_plots:3d}%
    \end{subfigure}
    \begin{subfigure}[t]{0.48\linewidth}
        \includegraphics[width=\linewidth]{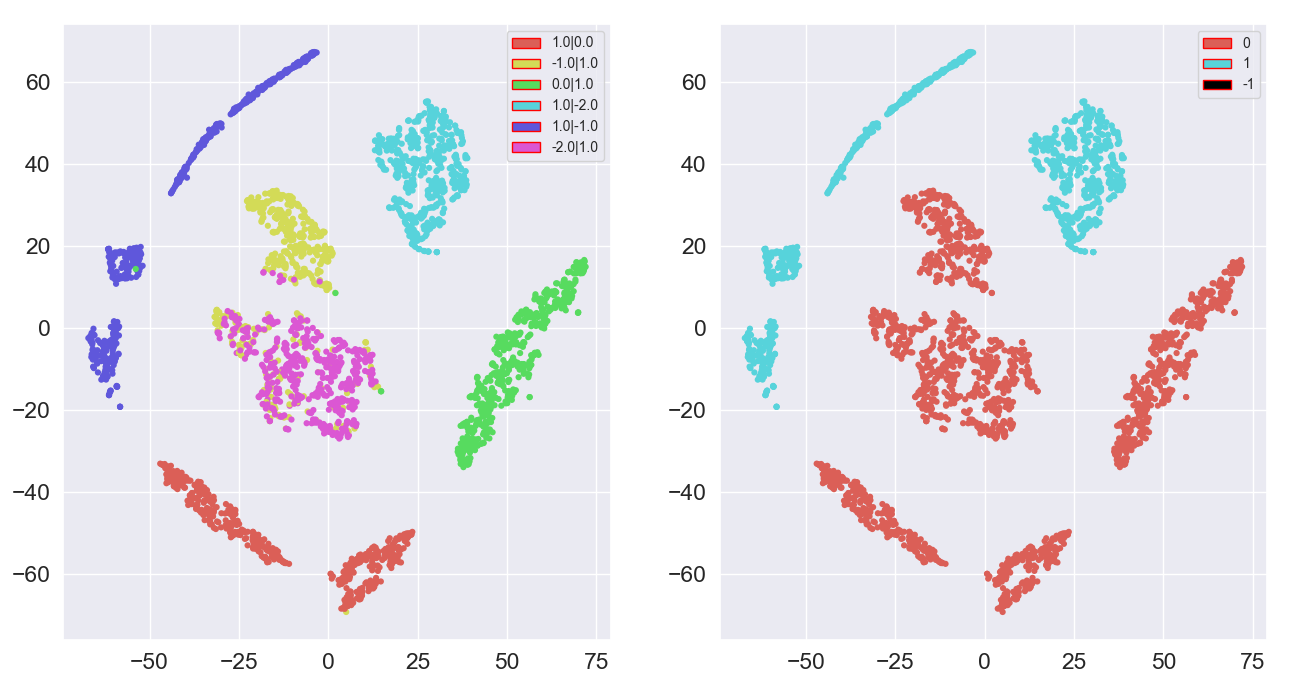}
        \caption{4-dimensional}
        \label{fig:feature_vector_plots:4d}%
    \end{subfigure}\\
    \begin{subfigure}[t]{0.48\linewidth}
        \includegraphics[width=\linewidth]{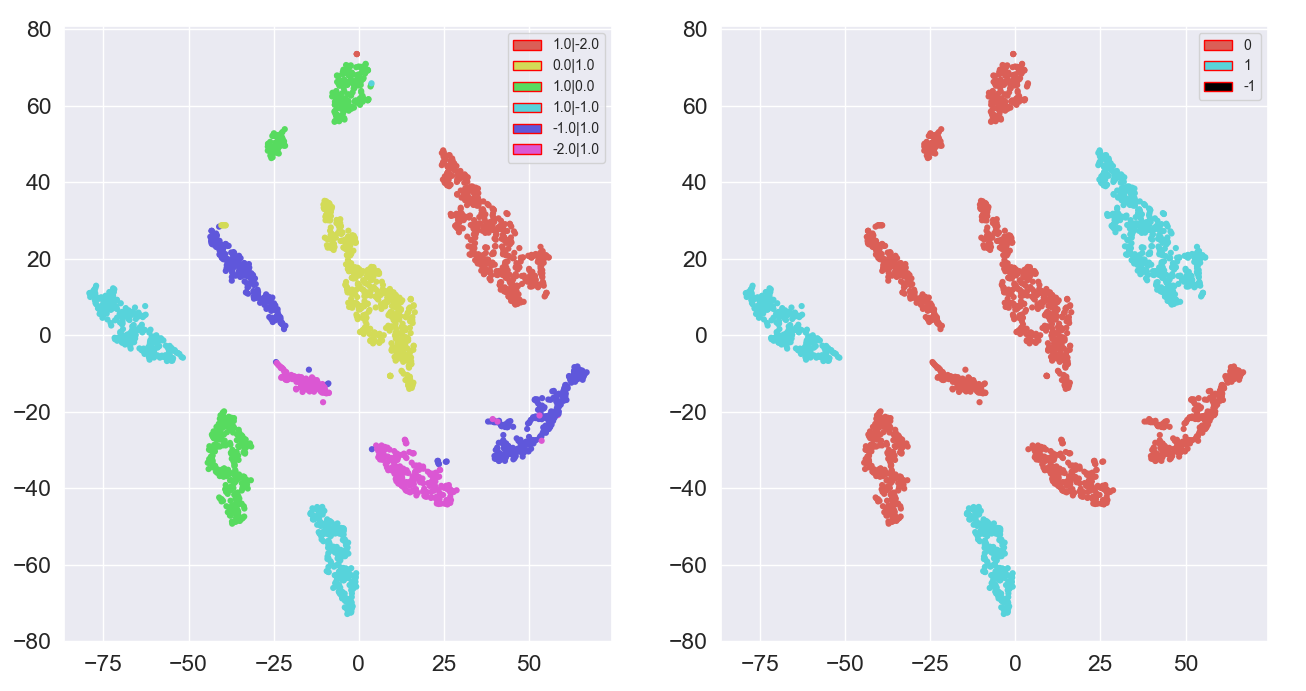}
        \caption{10-dimensional}
        \label{fig:feature_vector_plots:10d}%
    \end{subfigure}%
    \begin{subfigure}[t]{0.48\linewidth}
        \includegraphics[width=\linewidth]{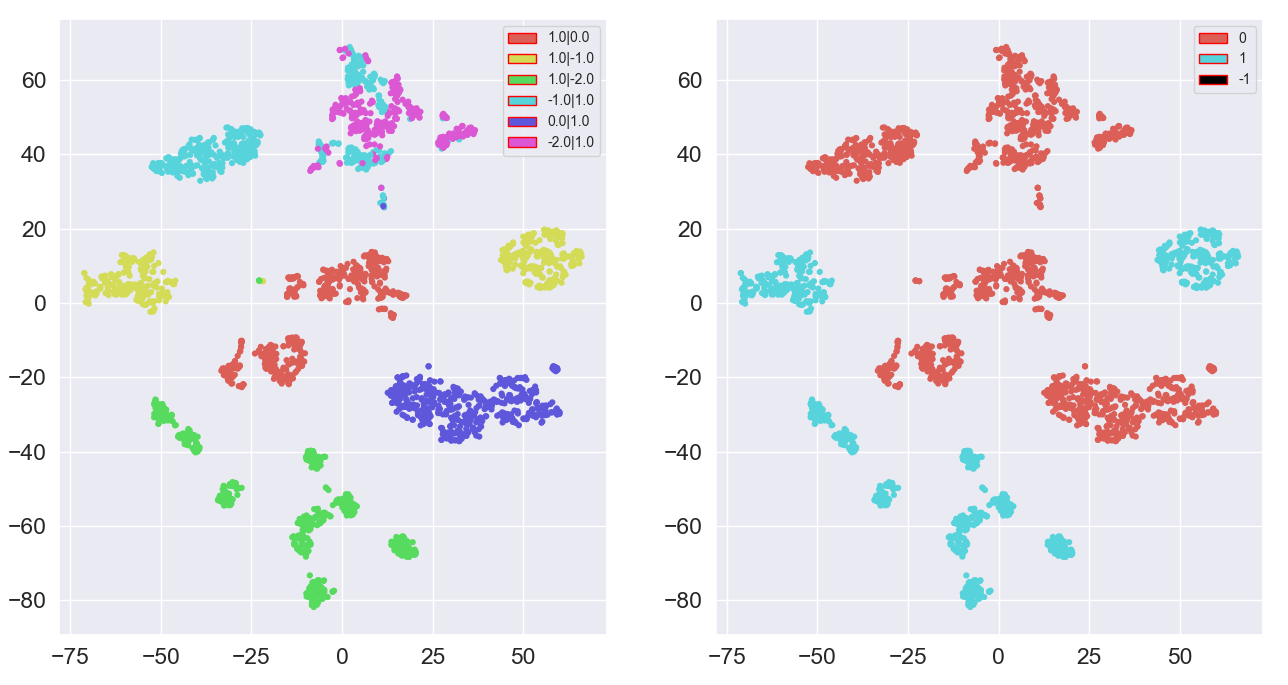}
        \caption{100-dimensional}
        \label{fig:feature_vector_plots:100d}%
    \end{subfigure}%
    \caption{Representation of the clusters learned by \ipdistill for Coin Game. Each point is a t-SNE projection of the feature vector (in different dimensions) output by the \ipdistill network for an input sequence of states. For each of the sub-figures, the figure on the left is colored based on actual rewards obtained by each agent ($r_1|r_2$). The figure on the right is colored based on clusters as learned by \ipdistill. \ipdistill correctly identifies two types of trajectories, one for cooperation and the other for defection.}
    \label{fig:feature_vector_plots:all}
\end{figure*}

\textbf{The first component is the state sequence encoder} that takes as input a sequence of states (input size is $4\times{}4\times{}3\times{}3$, where $4\times{}3\times{}3$ is the dimension of the game state, and the first index in the state input represents the data channel where each channel encodes data from both all the different colored agents and coins) and outputs a fixed dimension feature representation. 
We encode each state in the sequence using a common trunk of $3$ convolution layers with \textit{relu} activations and kernel-size $3\times{}3$, followed by a fully-connected layer with $100$ neurons to obtain a finite-dimensional feature representation.
This unified feature vector, called the trajectory embedding, is then given as input to the different prediction branches of the network.
We also experiment with different dimensions of this embedding and provide results in Figure~\ref{fig:feature_vector_plots:all}.

The two branches, which predict the self-reward and the opponent-reward (as shown in Figure~\ref{fig:high_level_approach}), independently use this trajectory embedding as input to compute appropriate output.
These branches take as input the trajectory embedding and use a dense hidden layer (with 100 neurons) with linear activation to predict the output.
We use the \textit{mean-squared error (MSE)} loss for the regression tasks in the prediction branches.
Linear activation allows us to cluster the trajectory embeddings using a linear clustering algorithm, such as Agglomerative Clustering~\cite{friedman2001elements}. 
In general, we can choose the number of clusters based on our desired level of granularity in differentiating outcomes. 
In the games considered in this paper, agents broadly have two types of policies. 
Therefore, we fix the number of clusters to two.

We use the \textit{Adam}~\cite{kingma2014adam} optimizer with learning-rate of $3e-3$.
We also experiment with K-Means clustering in addition to Agglomerative Clustering, and it also gives similar results.
We provide additional results of the clusters obtained using \ipdistill in Appendix~\ref{appendix:experimental-setup}.
\textbf{The second component is the oracle network} that outputs an action given a state.
For each oracle network, we encode the input state using $3$ convolution layers with kernel-size $2\times{}2$ and \textit{relu} activation.
To predict the action, we use $3$ fully-connected layers with \textit{relu} activation and the cross-entropy loss.
We use \textit{L2} regularization, and \textit{Gradient Descent} with the \textit{Adam} optimizer (learning rate $1e-3$) for all our experiments.

\subsection{\ipdistill: Pseudo-Code}
\label{appendix:gamedistill:pseudo-code}
\begin{algorithm}
    \caption{Pseduo-code for $collect\_data$}
    \begin{algorithmic}[1]
        \STATE{\bfseries Input:} Game Environment $Env$, Minimum Samples $min\_samples=2000$, Batch Size $batch=100, look\_back=5$
        \STATE $env=Env.spawn(batch)$
        \STATE $state\_q=Queue(batch,look\_back)$
        \STATE $reward\_seq\_dict=dict()$
        \STATE $keep\_running=True$
        \WHILE {$keep\_running$}
            \STATE $actions=random(env.actions,~size=(batch, env.n\_agents))$
            \STATE $rewards,~moves,~states=env.step(actions)$
            \STATE $check=False$
            \FOR{$b$ {\bfseries in} range($batch$)}
                \IF{$state\_q[b].full()$}
                    \STATE $state\_q[b].pop()$
                \ENDIF
                \STATE $state\_q[b].put(states[b])$
                \STATE
                \STATE // \textit{for any non zero reward tuple}
                \IF{$abs(rewards).sum()>0$}
                    \STATE $reward\_tpl=tuple(rewards)$
                    \IF{$reward\_tpl$ {\bfseries not in} $reward\_seq\_dict$}
                        \STATE $reward\_seq\_dict[reward\_tpl]=[~]$
                    \ENDIF
                    \STATE $obs=state\_q[b].pop\_all()$
                    \STATE $reward\_seq\_dict[reward\_tpl].append(obs)$
                    \STATE $check=True$
                \ENDIF
            \ENDFOR
            \IF{$check$}
                \STATE $keys=env.get\_all\_possible\_reward\_tuples()$
                \STATE $count\_stop=0$
                \FOR{$k$ {\bfseries in} $keys$}
                    \IF{$len(reward\_seq\_dict[k])>min\_samples$}
                    \STATE $count\_stop+=1$
                    \ENDIF
                \ENDFOR
                \IF{$count\_stop>=len(keys)$}
                    \STATE $keep\_running=False$
                \ENDIF
            \ENDIF
        \ENDWHILE
        \STATE
        \STATE $train\_data=[~]$~~~~~~~~~~~~~~~~// \textit{Collect the final training data}
        \FOR{$rewards\_tup$ {\bfseries in} $reward\_seq\_dict.keys()$}
            \FOR{$traj$ {\bfseries in} $t\_data[rewards\_tup][~:min\_samples]$}
                \STATE // \textit{trajectory has shape [look\_back,h,w,c] and ``$rewards\_tup$'' is tuple of rewards of agents}
                \STATE $traj = traj.permute(1,2,3,0).reshape(h,w,-1)$
                \STATE $train\_data.append((traj,~rewards\_tup))$
            \ENDFOR
        \ENDFOR
        \STATE \textbf{return} {$shuffle(train\_data)$} 
    \end{algorithmic}
    \label{algorithm:pseudo_code:collect_data}
\end{algorithm}
\begin{algorithm}[H]
    \caption{Pseduo-code for $create\_network$}
    \begin{algorithmic}[1]
        \STATE $net = conv(states\_placeholder,~kernel=3,~num\_outputs=64,~activation=relu)$
        \STATE $net = conv(net,~kernel=3,~num\_outputs=64,~activation=relu)$
        \STATE $net = conv(net,~kernel=3,~num\_outputs=64,activation=relu)$
        \STATE $feat = flatten(net)$~~~~~// \textit{the trajectory embedding}
        \STATE
        
        \STATE $self\_ft,~opp\_ft = FC(feat,~num\_outputs=100),~FC(feat,~num\_outputs=100)$
        \STATE
        \STATE // \textit{Predict the opponent and the self rewards}
        \STATE $s\_reward\_pred = FC(self\_ft,~num\_outputs=1)$
        \STATE $o\_reward\_pred = FC(opp\_ft,~num\_outputs=1)$
        \STATE \textbf{return} {$s\_reward\_pred,~o\_reward\_pred$} 
    \end{algorithmic}
    \label{algorithm:pseudo_code:create_network}
\end{algorithm}
\begin{algorithm}[H]
    \caption{Pseduo-code for $train\_network$}
    \begin{algorithmic}[1]
        \STATE{\bfseries Input:} Reward Prediction Network as $net$, data $train\_data$, loss term weights $\mathcal{A}$ and $\mathcal{B}$
        \STATE $optimizer = Adam(lr=0.003)$
        \WHILE{convergence}
            \STATE $state,~reward = sample(train\_data)$
            \STATE $my\_reward,~opp\_reward = net.forward(state)$
            \STATE $loss = \mathcal{A} * l2\_loss(my\_reward,~reward(0))~+~\mathcal{B} * l2\_loss(opp\_reward,~reward(1))$
            \STATE $loss.backward(),~~optimizer.step()$
        \ENDWHILE
    \end{algorithmic}
    \label{algorithm:pseudo_code:train_network}
\end{algorithm}
\begin{algorithm}[H]
    \caption{Pseduo-code for $create\_oracle\_net$}
    \begin{algorithmic}[1]
        \STATE{\bfseries Input:} Game Environment $env$
        \STATE $net = conv(state\_placeholder,~kernel=2,~num\_outputs=128, activation=relu)$
        \STATE $net = conv(net,~kernel=2,~num\_outputs=128, activation=relu)$
        \STATE $net = conv(net,~kernel=2,~num\_outputs=64, activation=relu)$
        \STATE $net = flatten(net)$
        \STATE $net = fc(net,~num\_outputs=128, activation=relu)$~~~// \textit{Encode the environment state}
        \STATE        
        \STATE // \textit{Predict the probability of taking an action}
        \STATE $logits = fc(net,~num\_outputs=env.num\_actions)$
        \STATE $action\_predict = softmax(logits)$
        \STATE{\bfseries return} Action predictions $action\_predict$ 
    \end{algorithmic}
    \label{algorithm:pseudo_code:create_oracle_net}
\end{algorithm}
\begin{algorithm}[H]
    \caption{Pseduo-code for $train\_oracle$}
    \begin{algorithmic}[1]
        \STATE{\bfseries Input:} Oracle Network $net$, Clustered trajectory data $train\_data$
        \STATE $actions\_data=[~]$
        \FOR{$data$ {\bfseries in} $train\_data$}
            \STATE$states = data[0]$
            \FOR{$i$ {\bfseries in} $range(1, len(states))$}
                \STATE $action=deduce\_move(states[i-1],~states[i])$
                \STATE $actions\_data.append((states[i-1],~action))$
            \ENDFOR
        \ENDFOR
        \STATE $optimizer=SGD(lr=0.01)$
        \WHILE {not convergence}
            \STATE$state,~action = sample(actions\_data)$
            \STATE$my\_action = net.forward(state)$
            \STATE$loss = cross\_entropy(my\_action,~action)$
            \STATE $loss.backward(),~optimizer.step()$
        \ENDWHILE
    \end{algorithmic}
    \label{algorithm:pseudo_code:train_oracles}
\end{algorithm}

\section{$SQLoss$: Emergence of Cooperation}
\label{appendix:sqloss_theory}
Equation~\ref{eq:pg_original} (Section~\ref{sec:approach:SQLoss:formulation}) describes the gradient for standard policy gradient. 
It has two terms.
The $log\pi^{1} (u_{t}^{1}|s_{t})$ term maximises the likelihood of reproducing the training trajectories $[(s_{t-1},u_{t-1},r_{t-1}),(s_{t},u_{t},r_{t}),(s_{t+1},u_{t+1},r_{t+1}),\dots]$.
The return term pulls down trajectories that have poor return.
The overall effect is to reproduce trajectories that have high returns. 
We refer to this standard loss as $Loss$ for the following discussion.
\begin{lemma}
    \label{lemma:lemma1}
    For agents trained with random exploration in the IPD, $Q_{\pi}(D|s_t) > Q_{\pi}(C|s_t)$ for all $s_t$.
\end{lemma}
Let $Q_{\pi}(a_t|s_t)$ denote the expected return of taking $a_t$ in $s_t$.
Let $V_{\pi}(s_t)$ denote the expected return from state $s_t$.
\begin{equation}
    \begin{split}
        Q_{\pi}(C|CC) &= 0.5*\big((-1) \; + \; V_{\pi}(CC)\big) \; + \; 0.5*\big((-3) + V_{\pi}(CD)\big)\\
        Q_{\pi}(C|CC) &= -2 \; + \; 0.5*\big(V_{\pi}(CC) + V_{\pi}(CD)\big)\\
        Q_{\pi}(D|CC) &= -1 \; + \; 0.5*\big(V_{\pi}(DC) + V_{\pi}(DD)\big)\\
        Q_{\pi}(C|CD) &= -2 \; + \; 0.5*\big(V_{\pi}(CC) + V_{\pi}(CD)\big)\\
        Q_{\pi}(D|CD) &= -1 \; + \; 0.5*\big(V_{\pi}(DC) + V_{\pi}(DD)\big)\\
        Q_{\pi}(C|DC) &= -2 \; + \; 0.5*\big(V_{\pi}(CC) + V_{\pi}(CD)\big)\\
        Q_{\pi}(D|DC) &= -1 \; + \; 0.5*\big(V_{\pi}(DC) + V_{\pi}(DD)\big)\\
        Q_{\pi}(C|DD) &= -2 \; + \; 0.5*\big(V_{\pi}(CC) + V_{\pi}(CD)\big)\\
        Q_{\pi}(D|DD) &= -1 \; + \; 0.5*\big(V_{\pi}(DC) + V_{\pi}(DD)\big)
    \end{split}
\end{equation}
Since $V_{\pi}(CC)=V_{\pi}(CD)=V_{\pi}(DC)=V_{\pi}(DD)$ for randomly playing agents, $Q_{\pi}(D|s_t) > Q_{\pi}(C|s_t)$ for all $s_t$.

\begin{lemma}
\label{lemma:lemma2}
Agents trained to only maximize the expected reward in IPD will converge to mutual defection. 
\end{lemma}
This lemma follows from Lemma~\ref{lemma:lemma1}.
Agents initially collect trajectories from random exploration. 
They use these trajectories to learn a policy that optimizes for a long-term return. 
These learned policies always play $D$ as described in Lemma~\ref{lemma:lemma1}.

Equation 7 describes the gradient for $SQLoss$. 
The $log\pi^{1} (u_{t-1}^{1}|s_{t})$ term maximises the likelihood of taking $u_{t-1}$ in $s_{t}$.
The imagined episode return term pulls down trajectories that have poor imagined return. 

\begin{lemma}
\label{lemma:lemma3}
Agents trained on random trajectories using only $SQLoss$ oscillate between $CC$ and $DD$. 
\end{lemma}

For IPD, $s_{t} = (u_{t-1}^{1}, u_{t-1}^{2})$.
The $SQLoss$ maximises the likelihood of taking $u_{t-1}$ in $s_{t}$ when the return of the imagined trajectory $\hat R_{t}(\hat \uptau_1)$ is high.

Consider state $CC$, with $u_{t-1}^{1}=C$.
$\pi^{1} (D|CC)$ is randomly initialised. 
The $SQLoss$ term reduces the likelihood of $\pi^{1} (C|CC)$ because $\hat R_{t}(\hat \uptau_1) < 0$. 
Therefore, $\pi^{1} (D|CC) > \pi^{1} (C|CC)$.

Similarly, for $CD$, the $SQLoss$ term reduces the likelihood of $\pi^{1} (C|CD)$.
Therefore, $\pi^{1} (D|CD) > \pi^{1} (C|CD)$.
For $DC$, $\hat R_{t}(\hat \uptau_1) = 0$, therefore $\pi^{1} (D|DC) > \pi^{1} (C|DC)$.
Interestingly, for $DD$, the $SQLoss$ term reduces the likelihood of $\pi^{1} (D|DD)$ and therefore $\pi^{1} (C|DD) > \pi^{1} (D|DD)$. 

Now, if $s_t$ is $CC$ or $DD$, then $s_{t+1}$ is $DD$ or $CC$ and these states oscillate. 
If $s_t$ is $CD$ or $DC$, then $s_{t+1}$ is $DD$, $s_{t+2}$ is $CC$ and again $CC$ and $DD$ oscillate. 
This oscillation is key to the emergence of cooperation as explained in section 2.3.1.

\begin{lemma}
\label{lemma:lemma4}
For agents trained using both standard loss and $SQLoss$, $\pi (C|CC) > \pi^{1} (D|CC)$.
\end{lemma}
For $CD$, $DC$, both the standard loss and $SQLoss$ push the policy towards $D$.
For $DD$, with sufficiently high $\kappa$, the $SQLoss$ term overcomes the standard loss and pushes the agent towards $C$.
For $CC$, initially, both the standard loss and $SQLoss$ push the policy towards $D$.
However, as training progresses, the incidence of $CD$ and $DC$ diminish because of $SQLoss$ as described in Lemma 3.
Therefore, $V_{\pi}(CD) \approx V_{\pi}(DC)$ since agents immediately move from both states to $DD$. 
Intuitively, agents lose the opportunity to exploit the other agent. 
In equation 9, with $V_{\pi}(CD) \approx V_{\pi}(DC)$, $Q_{\pi}(C|CC) > Q_{\pi}(D|CC)$ and the standard loss pushes the policy so that $\pi (C|CC) > \pi (D|CC)$. 
This depends on the value of $\kappa$.
For very low values, the standard loss overcomes $SQLoss$ and agents defect.
For very high values, $SQLoss$ overcomes standard loss, and agents oscillate between cooperation and defection.
For moderate values of $\kappa$ (as shown in our experiments), the two loss terms work together so that $\pi (C|CC) > \pi (D|CC)$.

\section{Games with more than 2 players}
\label{appendix:beyond_2_players}
Our formulation of $SQLoss$ has the distinct advantage of being fully ego-centric, that is, the agent that is learning does not require any information regarding its opponents. This feature enables a straightforward extension of $SQLoss$ beyond the two agent setting, without any change in each agent's learning algorithm. In order to test this extension of $SQLoss$ beyond 2-players, we consider as an example, the problem described in the popular Braess' paradox, which is a well-known extension of the Prisoner's Dilemma problem to more than 2 agents.
\begin{figure}[htb!]
    \begin{minipage}{0.48\textwidth}
        \centering
        \includegraphics[page=2,width=0.99\linewidth]{images/braess_paradox.pdf}
        \caption{Braess' Paradox}
        \label{fig:braess_paradox}
    \end{minipage}\hfill
    \begin{minipage}{0.48\textwidth}
        \centering
        \begin{subfigure}[t]{0.99\linewidth}
            \includegraphics[width=\linewidth]{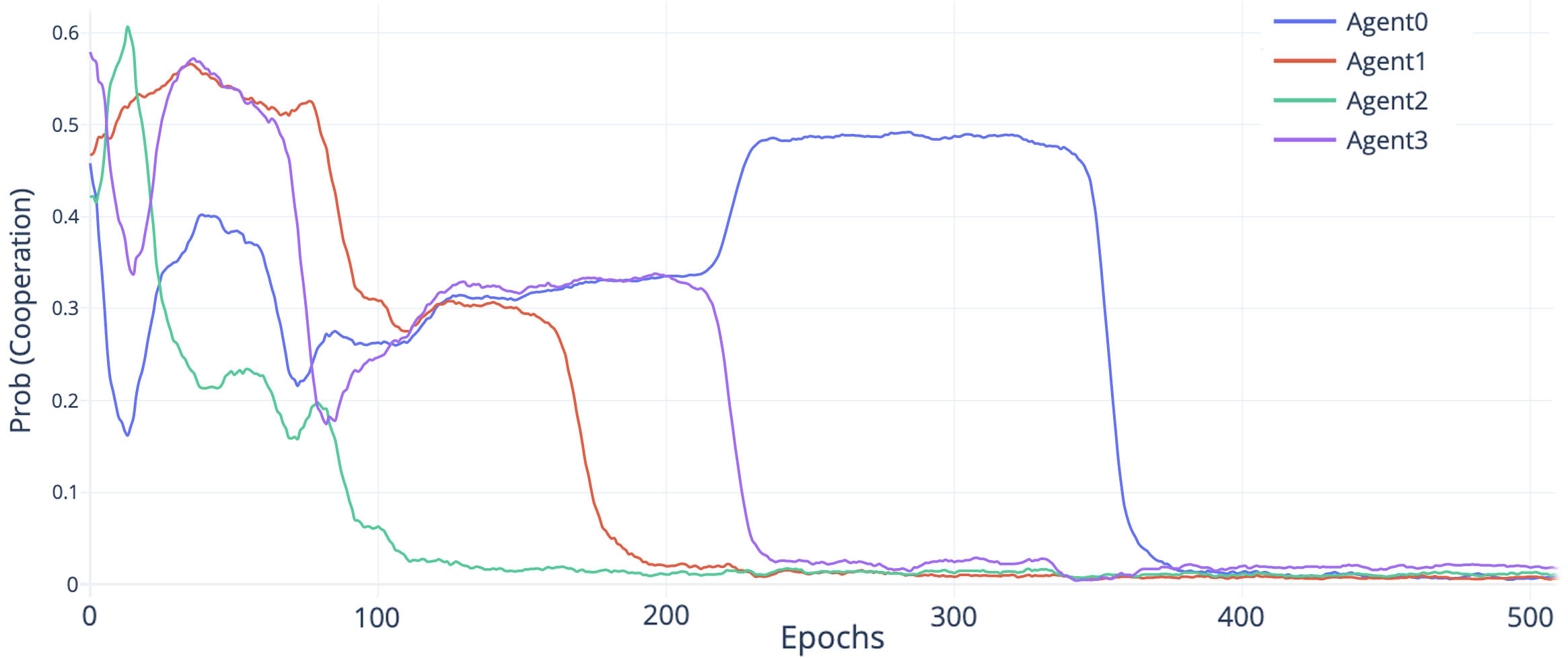}
            \caption{Results for 4 NL agents in Braess' Environment}
            \label{fig:braess_4player_nlpg}
        \end{subfigure}~~\\
        \begin{subfigure}[t]{0.99\linewidth}
            \includegraphics[width=\linewidth]{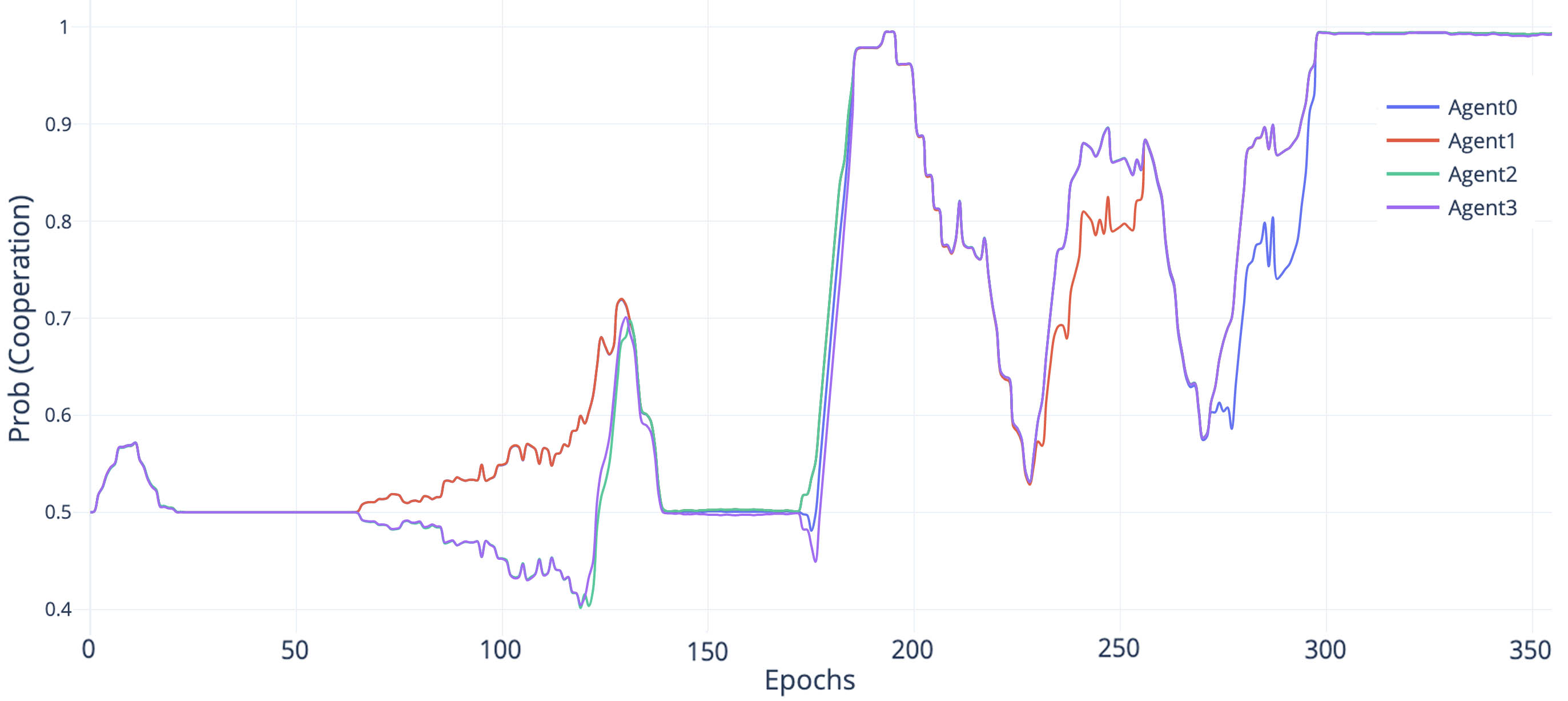}
            \caption{Results for 4 SQLearner agents in Braess' Environment}
            \label{fig:braess_4player_sqpg}%
        \end{subfigure}%
        \caption{%
            \label{fig:braess_results}%
        }%
    \end{minipage}
\end{figure}
We construct a simplified environment to interpret the policies in the problem given in Braess' paradox as a sequential social dilemma problem. This problem, which we henceforth refer to as the Braess' problem, is concerned with traffic flow from a source to a destination with two alternative paths, through two intermediate locations. This is illustrated in Fig.~\ref{fig:braess_paradox}.

In this, each agent has to travel from the source (Start) to the destination (End) by choosing one of these paths. Each path has two segments separated by the respective intermediate location and the cost of traversing each segment (proportional to the time of travel) is either fixed or is determined based on the number of agents using that segment. This cost function is shown in Eq.~\ref{eq:braess_cost_function}. In the original setup, the two intermediate locations are not connected and each agent has to only choose between one of these paths. The equilibrium solution in this case is for half the agents to choose one path and the remaining half the other.  The modification that entails is the paradox is connecting the two intermediate locations with a cost free directed bridge, thus leading a third possible path for all agents. This modification shifts the equilibrium in such a way that all agents now prefer to choose the new path that uses this directed bridge, although this results in a per-agent cost that is higher than each agent's original cost in the absence of this new path. This problem has been well studied and the paradox highlights the fact that in strategic settings, increase in the number of available options for all agents, may also lead to decrease in utility for all agents individually and collectively.

We model this problem as a sequential social dilemma, where we define the initial strategy of the agents to choose one of the two original paths as \textit{Cooperation} and the strategy that results in an agent choosing the new path with the directed bridge as \textit{Defection}. For simplicity of exposition, we assign integer IDs to agents and define Cooperation for agents with odd-IDs as choosing the path Start-A-End in Fig.~\ref{fig:braess_paradox} and Cooperation for agents with even-IDs as choosing the path Start-B-End in Fig~\ref{fig:braess_paradox}. In this notation, the Defection can be defined as an agent choosing the path Start-A-B-End. In the Fig.~\ref{fig:braess_paradox} we provide reward numbers, such that the paradox is realized.

As described earlier, in this game, Cooperate and Defect have the following interpretation:
\begin{enumerate}
    \item For an odd numbered vehicle – Cooperation implies taking the Start-A-End route.
    \item For an even numbered vehicle – Cooperation implies taking the Start-B-End route (This and above point follow from the assumption that before addition of the new edge A-B, the vehicles were in this equilibrium).
    \item For any vehicle – Defection implies taking the Start-A-B-End route.
\end{enumerate}
Let $n_{X-Y}$ denote the number of agents using the edge $X-Y$. Then the reward structure for the odd-numbered vehicles (denoted by $R_{odd}$) and the even-numbered vehicles (denoted by $R_{even}$) is computed as shown below.
\begin{equation}%
    \begin{split}
        N_{0} &= \text{Total no. of agents}\\
        R_{0} &= \text{base reward for the agents} = (2.5 * N_{0})/2\\
        R_{odd} &=
        \begin{cases}
          -(n_{Start-A} + n_{B-End}) & \text{if}\ \textit{Defection} \\
          -(n_{Start-A} + R_{0}), & \text{if}\ \textit{Cooperation} \\
        \end{cases}\\
        R_{even} &=
        \begin{cases}
          -(n_{Start-A} + n_{B-End}) & \text{if}\ \textit{Defection} \\
          -(R_{0} + n_{B-End}), & \text{if}\ \textit{Cooperation} \\
        \end{cases}
    \end{split}
    \label{eq:braess_cost_function}
\end{equation}
We performed two separate set of experiments in this game with 4 and 6 agents. For each setup, we simulated the result when all agents are selfish learners (the $SL$ agent(s)) and also when all agents use $SQLoss$ (the $SQLearner$(s)). For both these setups, as expected, we observe that when using selfish learners, all agents converge to Defection and when using $SQLoss$, all agents converge to Cooperation. We present the results for the Braess' Paradox experiment in Figure~\ref{fig:braess_results}.

The Figure~\ref{fig:braess_4player_nlpg} shows the results for 4 SL-agents playing in the environment. From the figure it is clear that if 4 SL agents play in the environment, then they eventually end up defecting i.e. $\lim_{E \to \infty}\mathbb{P}(Cooperation) \to 0$, where $E$ denotes the epoch. Figure~\ref{fig:braess_4player_sqpg} illustrates the behaviour of the $SQLearner$ agents. The SQLearner agents eventually learn to cooperate and thus the $\lim_{E \to \infty}\mathbb{P}(Cooperation) \to 1$. To the best of our knowledge, this is the first demonstration of selfish agents learning to cooperate in a sequential social dilemma with more than 2 agents.

\section{Experimental Details}
\label{appendix:experimental-setup}

\subsection{Infrastructure for Experiments}
\label{appendix:experimental-setup:computation-req}
We performed all our experiments on an AWS instance with the following specifications. We use a 64-bit machine with Intel(R) Xeon(R) Platinum 8275CL CPU @ 3.00GHz installed with Ubuntu 16.04LTS operating system. It had a RAM of 189GB and 96 CPU cores with two threads per core. We use the TensorFlow framework for our implementation.

\subsection{SQLoss}
\label{appendix:experimental-setup:sqloss}
For our experiments with the Selfish and Status-Quo Aware Learner ($SQLearner$), we use policy gradient-based learning to train an agent with the Actor-Critic method~\cite{sutton2011reinforcement}. Each agent is parameterized with a policy actor and critic for variance reduction in policy updates. During training, we use $\alpha=1.0$ for the REINFORCE and $\beta=0.5$ for the imaginative game-play. We use gradient descent with step size, $\delta = 0.005$ for the actor and $\delta = 1$ for the critic. We use a batch size of $4000$ for Lola-PG~\cite{foerster2018learning} and use the results from the original paper. We use a batch size of 200 for $SQLearner$ for roll-outs and an episode length of 200 for all iterated matrix games. We use a discount rate ($\gamma$) of $0.96$ for the Iterated Prisoners' Dilemma, Iterated Stag Hunt, and Coin Game. For the Iterated Matching Pennies, we use $\gamma = 0.9$ to be consistent with earlier works. The high value of $\gamma$ allows for long time horizons, thereby incentivizing long-term rewards. Each agent randomly samples $\kappa$ from $\mathbb{U} \in (1, z)$ ($z=10$, discussed in Appendix~\ref{appendix:effect-z-convergence}) at each step.

\section{Visualizing clusters obtained from \ipdistill}
\label{appendix:gamedistill-clustering}
\begin{figure}[h!]
    \centering
    \includegraphics[width=0.85\linewidth]{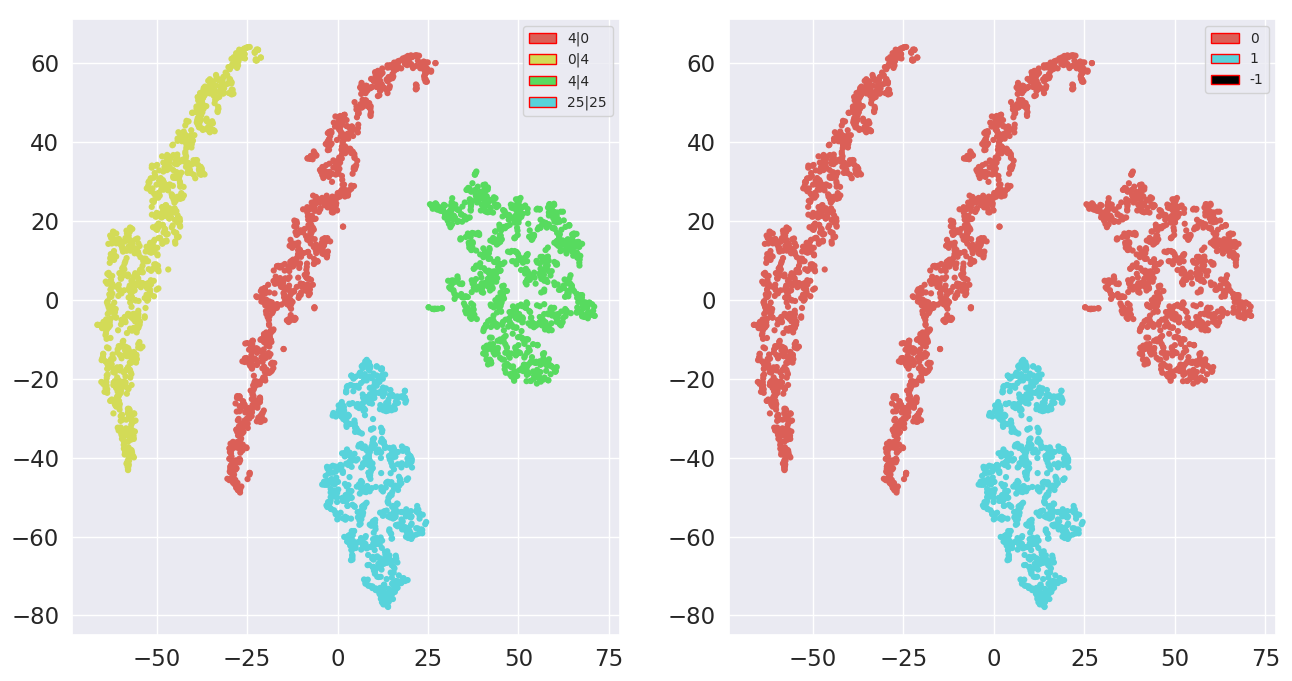}
    \caption{t-SNE plot for the trajectory embeddings obtained for the Stag Hunt game using \ipdistill along with the identified cooperation and defection clusters. Details on how to read the figures is provide with Figure~\ref{fig:feature_vector_plots:all}}
    \label{fig:stag_hunt_cluster_plot}%
\end{figure}

Figures~\ref{fig:feature_vector_plots:all} and ~\ref{fig:stag_hunt_cluster_plot} show the clusters obtained for the state sequence embedding for the Coin Game and the dynamic variant of Stag Hunt respectively.

In the figures, each point is a t-SNE projection of the feature vector (in different dimensions) output by the \ipdistill network for an input sequence of states. For each of the sub-figures, the figure on the left is colored based on actual rewards obtained by each agent ($r_1|r_2$). The figure on the right is colored based on clusters, as learned by \ipdistill. \ipdistill correctly identifies two types of trajectories, one for cooperation and the other for defection for both the games Coin Game and Stag-Hunt.

Figure~\ref{fig:feature_vector_plots:all} also shows the clustering results for different dimensions of the state sequence embedding for the Coin Game. We observe that changing the size of the embedding does not have any effect on the results.

\section{Results for Visual Stag-Hunt game with \ipdistill and $SQLoss$}
\label{appendix:results:visual-staghunt-with-gamedistill-and-sqloss}
Figure~\ref{fig:staghunt_game_results_with_gamedistill_and_sqloss} shows the results for the $SQLearner$ agents using the trained oracles obtained from \ipdistill along with $SQLoss$ in the visual StagHunt environment.


\section{Results for the Iterated Chicken Game (ICG) using $SQLoss$}
\label{appendix:results:chicken}
\begin{figure}[htb!]
    \begin{minipage}{0.46\textwidth}
        \centering
        \includegraphics[width=\linewidth]{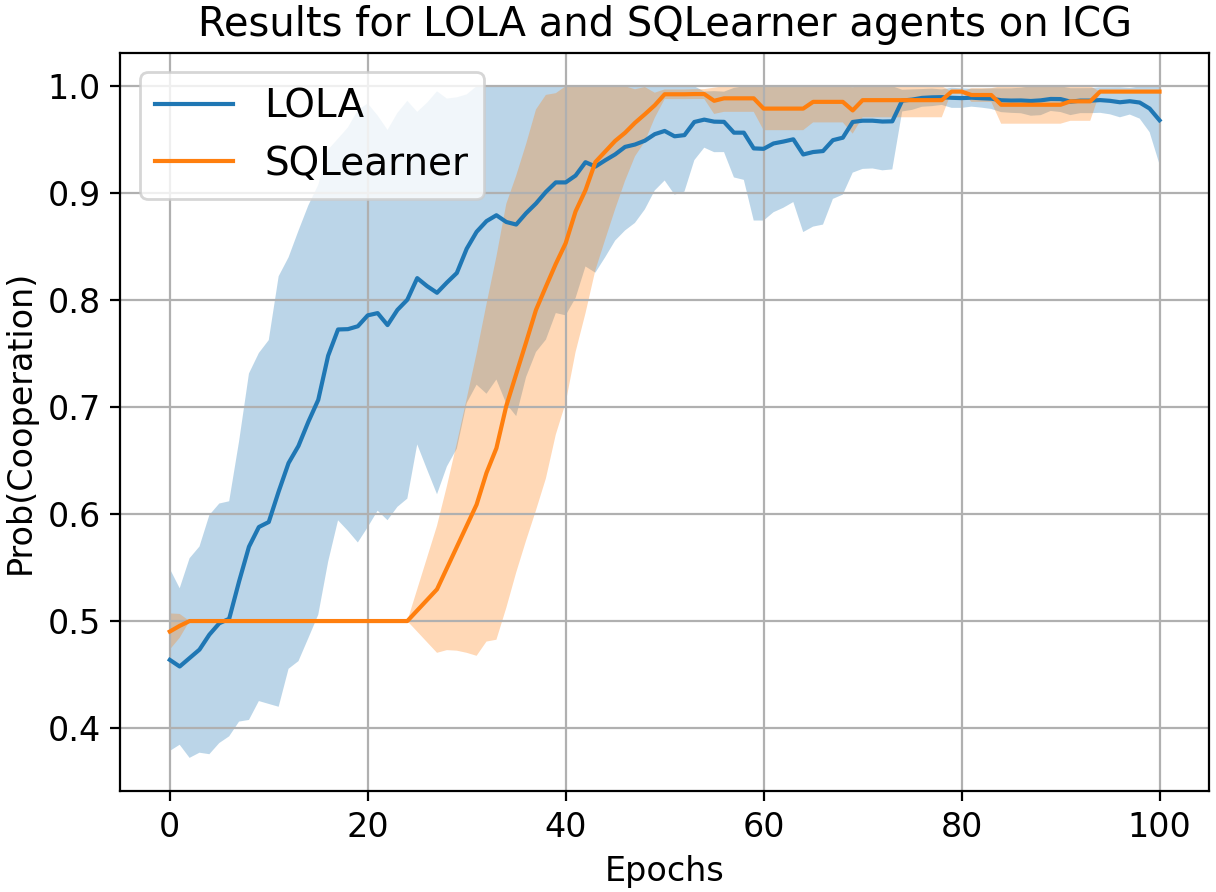}
        \caption{P(Cooperation) for LOLA and $SQLearner$ agents in ICG. Both agents eventually obtain a near optimal probability of 1.0}
        \label{fig:chicken_game_results}%
    \end{minipage}\hfill%
    \begin{minipage}{0.48\textwidth}
        \centering
        \includegraphics[width=\linewidth]{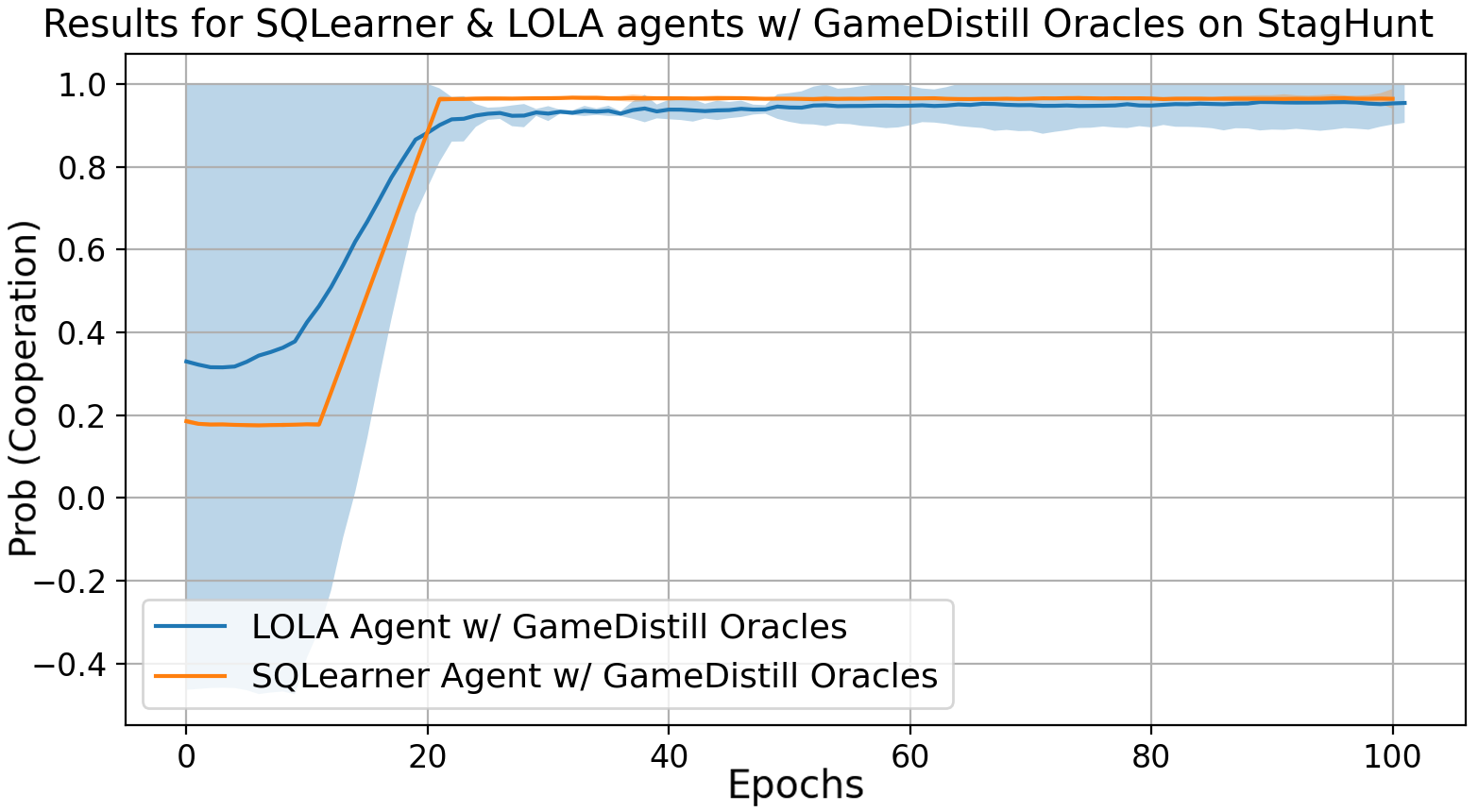}
        \caption{P(Cooperation) for $SQLearner$ agents in visual StagHunt with \ipdistill oracles. Both agents eventually learn to capture Stag roughly 95\% of the time which is close to the optimal for the game.}
        \label{fig:staghunt_game_results_with_gamedistill_and_sqloss}%
    \end{minipage}\hfill
\end{figure}

Figure~\ref{tab:payoff_matrix_chicken_game} presents the payoff matrix for the Chicken Game. In Figure~\ref{fig:chicken_game_results} we compare the results of a $SQLearner$ agent on the ICG Game with a LOLA agent. The payoff matrix for the game is shown in the Table~\ref{tab:payoff_matrix_chicken_game}.
\begin{table}[hbt!]
    \centering
    \renewcommand{\arraystretch}{1.2}
    \begin{tabular}{c|c|c}
         & $C$ & $D$ \\ \hline
         $C$ & (-1, -1) & (-3, 0) \\ \hline
         $D$ & (0, -3) & (-4, -4) \\ \hline
    \end{tabular}
    \caption{\label{tab:payoff_matrix_chicken_game}%
    Chicken Game}
\end{table}%
From the payoff, it is clear that the agents may defect out of greed. In this game also, $SQLearner$ agents coordinate successfully to obtain a near-optimal NDR value ($0$) for this game.

\section{$SQLoss$ vs \ipdistill}
\label{appendix:sqloss_vs_gamedistill}
In this section we present the results for the additional experiments we do to disentangle the effect of \ipdistill and $SQLoss$ for the performance in Figure~\ref{fig:coingame_results} by training a variant of LOLA that uses the oracles learned by \ipdistill.

To discuss the observation regarding how much of the performance shown in Figure~\ref{fig:coingame_results} is attributable to \ipdistill, we further discuss Figures~\ref{fig:results_IPD} and \ref{fig:coingame_results}.
Figure~\ref{fig:results_IPD} shows the different methods applied to the matrix formulation of the Iterated Prisoner's Dilemma (IPD) game.
In this formulation, each method ($SQLearner$, LOLA, etc.) uses the same reduced action space consisting of cooperation and defection.
Figure~\ref{fig:results_IPD} shows how agents trained using $SQLoss$ outperform those trained using LOLA in this setting.
For our experiments on the Coin Game (Figure~\ref{fig:coingame_results}), we use (\ipdistill + $SQLoss$) against LOLA.

From the results above, it is clear that the performance of the $SQLearner$ agent is due to the $SQLoss$ which promotes cooperation (as shown in multiple matrix games), the \ipdistill algorithm further expands this gain by reducing the dimensionality of the games with visual input.
The other baselines, e.g. Lola-PG and $SL$ do not use \ipdistill. To further test this hypothesis, we use the oracles developed using \ipdistill with the $SL$ agents, and observe that the $SL$ agents with \ipdistill do converge faster to \textit{DD} than traditional $SL$ agents, motivating us to use \ipdistill.

\section{Illustrations of Trained Oracle Networks}
\label{appendix:oracle:illustration}
\subsection{Coin Game}
\begin{figure}[h]
    \centering
    \includegraphics[width=0.85\linewidth]{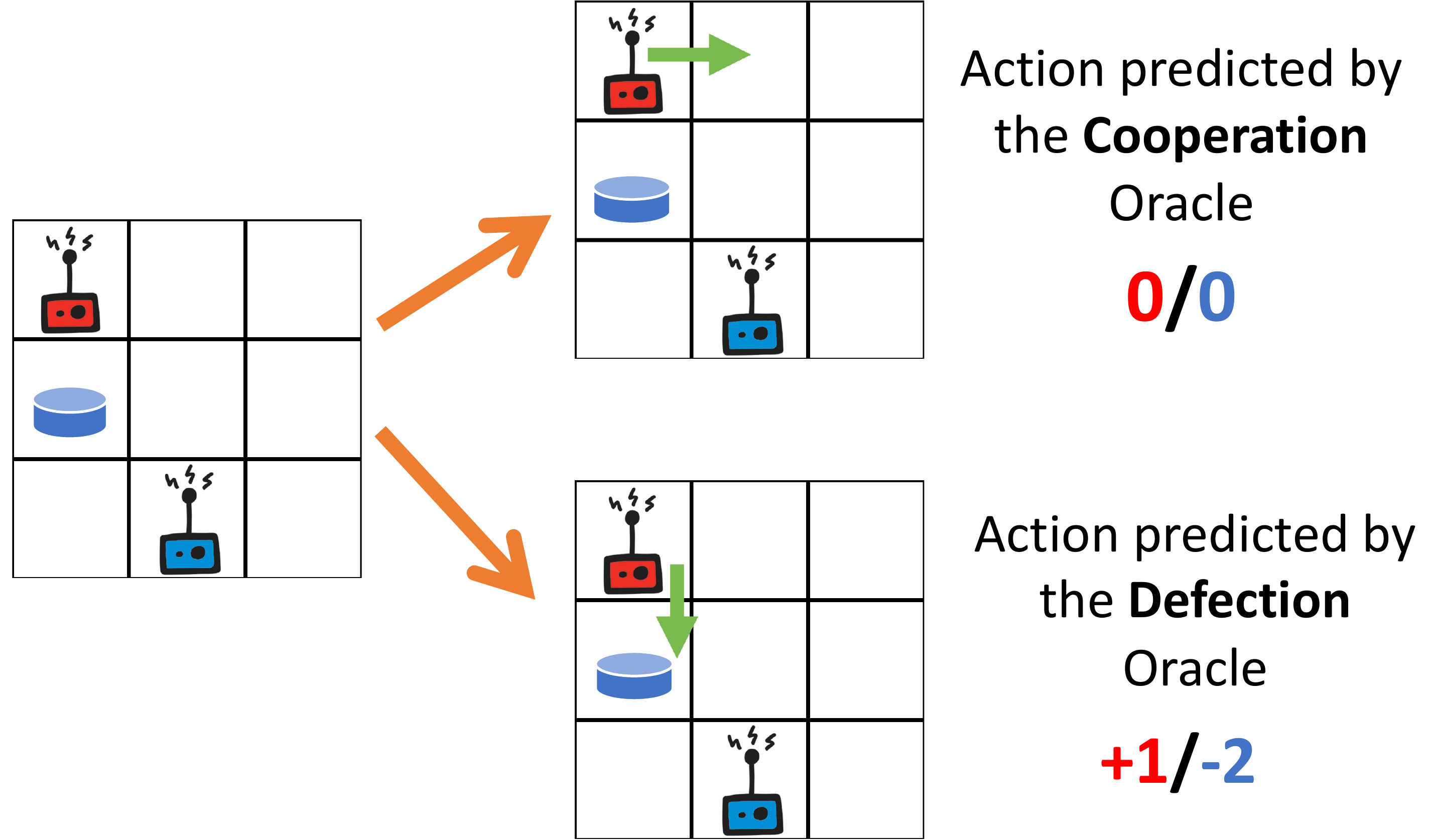}
    \caption{Illustrative predictions of the oracle networks learned by the Red agent using \ipdistill in the Coin Game. The numbers in \textcolor{red}{red}/\textcolor{blue}{blue} show the rewards obtained by the Red and the Blue agent respectively. The cooperation oracle suggests an action that avoids picking the coin of the other agent while the defection oracle suggests an action that picks the coin of the other agent}
    \label{fig:oracle_predictions}
\end{figure}

Figure~\ref{fig:oracle_predictions} shows the predictions of the oracle networks learned by the Red agent using \ipdistill in the Coin Game.
We see that the cooperation oracle suggests an action that avoids picking the coin of the other agent (the Blue coin).
Analogously, the defection oracle suggests a selfish action that picks the coin of the other agent.
Empirically, we train the Cooperation and Defection oracles and obtain a probability of picking self-colored coin (or $P(Cooperation)$) close to $0.916$ and $0.006$ respectively. 

\subsection{StagHunt}
\ipdistill produces two oracles when trained on the StagHunt environment. One of the trained oracles results in both the agents capturing the Stag 99\% of the times i.e. $Probability~(Capturing~a~Stag|Stag,~Hare) = 0.99$,  while the other oracle forces the agents to eat the Hare with similar probability of 0.99. Both the oracles learn to accurately capture the desired item in the environment.

\section{$SQLoss$: Effect of $z$ on convergence to cooperation}
\label{appendix:effect-z-convergence}
We explore the effect of the hyper-parameter $z$ (Section~\ref{sec:approach}) on convergence to cooperation, we also experiment with varying values of $z$.
In the experiment, to imagine the consequences of maintaining the status quo, each agent samples $\kappa_{t}$ from the Discrete Uniform distribution $\mathbb{U} \{1,z\}$.
A larger value of $z$ thus implies a larger value of $\kappa_{t}$ and longer imaginary episodes.
We find that larger $z$ (and hence $\kappa$) leads to faster cooperation between agents in the IPD and Coin Game. This effect plateaus for $z > 10$.
However varying and changing $\kappa_{t}$ across time also increases the variance in the gradients and thus affects the learning. We thus use $\kappa = 10$ for all our experiments.

\section{$SQLearner$: Exploitability and Adaptability}
\label{appendix:exploitability}
Given that an agent does not have any prior information about the other agent, it must learn its strategy based on its opponent's strategy. To evaluate an $SQLearner$ agent's ability to avoid exploitation by a selfish agent, we train one $SQLearner$ agent against an agent that always defects in the Coin Game. We find that the $SQLearner$ agent also learns to always defect. This persistent defection is important since given that the other agent is selfish, the $SQLearner$ agent can do no better than also be selfish. To evaluate an $SQLearner$ agent's ability to exploit a cooperative agent, we train one $SQLearner$ agent with an agent that always cooperates in the Coin Game. In this case, we find that the $SQLearner$ agent learns to always defect. This persistent defection is important since given that the other agent is cooperative, the $SQLearner$ agent obtains maximum reward by behaving selfishly. Hence, the $SQLearner$ agent is both resistant to exploitation and able to exploit, depending on the other agent's strategy.

\end{document}